\documentclass[journal,12pt,draftclsnofoot,onecolumn]{IEEEtran}
\usepackage{amsmath}
\usepackage{bm}
\usepackage{amssymb}
\usepackage{latexsym}
\usepackage{multirow}
\usepackage{epsfig}
\usepackage{graphics}
\usepackage{mathrsfs}
\usepackage{algorithm}
\usepackage{algpseudocode}
\usepackage{subcaption}
\usepackage{color}
\usepackage{graphicx}

\usepackage{algcompatible}
\usepackage{array}
\usepackage{amsfonts}
\usepackage{mdwmath}
\usepackage{mdwtab}
\usepackage{eqparbox}
\usepackage{epstopdf}

\newcommand{\bc}{\begin{center}}
\newcommand{\ec}{\end{center}}
\newcommand{\be}{\begin{equation}}
\newcommand{\ee}{\end{equation}}
\newcommand{\bea}{\begin{eqnarray}}
\newcommand{\eea}{\end{eqnarray}}

\newtheorem{theorem}{Theorem}[section]

\newtheorem{lemma}[theorem]{Lemma}

\begin{document}
\begin{titlepage}
\title{Robust Beamforming Design in a NOMA
Cognitive Radio Network Relying on SWIPT}

\author{Haijian Sun, \emph{Student Member, IEEE}, Fuhui Zhou, \emph{Member, IEEE}, Rose Qingyang Hu, \emph{Senior Member, IEEE}, and  Lajos Hanzo, \emph{Fellow, IEEE}

\thanks{Haijian Sun, Fuhui Zhou and Rose Qingyang Hu are with the Department of Electrical and Computer Engineering,
Utah State University, USA (e-mail: rosehu@ieee.org). Lajos Hanzo is with the University of Southampton, Southampton SO17 1BJ, U.K. (e-mail: lh@ecs.soton.ac.uk).

 This work has been accepted in part at the IEEE International Conference on Communications (ICC), Kansas City, MO, May 2018.}

}
\end{titlepage}
\maketitle

\begin{abstract}
This paper studies a multiple-input single-output non-orthogonal multiple access cognitive radio network relying on simultaneous wireless information and power transfer.  A realistic non-linear energy harvesting  model is applied and a power splitting architecture is adopted at each secondary user (SU). Since it is difficult to obtain perfect channel state information (CSI)  in practice, instead either a bounded or gaussian CSI error model is considered. Our robust beamforming and power splitting
ratio are jointly designed for two problems with different objectives, namely that of minimizing the transmission power of the cognitive base station and that of maximizing the total harvested energy of the SUs, respectively. The optimization problems are challenging to solve, mainly because of the non-linear structure of the energy harvesting and CSI  errors models.  We converted them into convex forms by using semi-definite relaxation. For the minimum transmission power problem, we obtain the rank-2 solution under the bounded CSI error model, while for the maximum energy harvesting problem, a two-loop procedure using a one-dimensional search is proposed. Our simulation results show that the proposed scheme significantly outperforms its traditional orthogonal multiple access counterpart. Furthermore, the performance using the gaussian CSI error model is  generally  better than that using  the bounded CSI error model.
\end{abstract}


\IEEEpeerreviewmaketitle
\newpage

\section{Introduction}
\IEEEPARstart{N}{on-orthogonal} multiple access (NOMA) has been recognized as one of the most promising techniques for next-generation wireless communication systems due to its capability of supporting a high spectral efficiency (SE) and massive connectivity \cite{Z. Ding}.  Since its design philosophy may be combined with diverse transceivers, it has drawn tremendous attention in multiple-antenna systems \cite{Z. Ding1}-\cite{sun1}, in cooperative networks \cite{Z. Ding2}, \cite{Yiran}, in device-to-device (D2D) networks \cite{sun2}, as well as in downlink and uplink multi-cell networks \cite{Z.Zhang}. In contrast to classic orthogonal multiple access (OMA), NOMA provides simultaneous access to multiple  users at the same time and  on the same frequency band, for example by using power-domain multiplexing.  In order to decrease the mutual interference among different users of power-domain NOMA, successive interference cancellation (SIC) may be applied by the receivers \cite{Z. Ding}. It has been shown that NOMA is capable of achieving a higher SE and energy efficiency (EE)  than OMA \cite{Z. Ding1}-\cite{Z.Zhang}.

As another promising technique of improving the SE, cognitive radio (CR) techniques have also been investigated for decades, where the secondary users (SUs) may  access the spectrum bands of the primary users (PUs), as long as the interference caused by SUs is tolerable \cite{F. H. Zhou}. According to \cite{A. Goldsmith}, in order to implement CR in practice, three operational models have been proposed, namely, opportunistic spectrum access, spectrum sharing,  and sensing-based enhanced spectrum sharing. It is envisioned that the combination of NOMA with CR is capable of further improving the SE. As a benefit of its low implementational complexity, spectrum sharing has been widely applied. In \cite{Y. Liu}-\cite{L. Lv1}, the authors analyzed the performance of a spectrum sharing CR combined with NOMA. It was shown that the SE can be significantly improved by using NOMA in CR compared to that achieved by using OMA in CR.

On the other hand,  the increasing greenhouse gas emissions  have become a major concern also in the design of wireless communication networks. According to \cite{X. Huang}, cellular networks world-wide  consume approximately 60 billion kWh energy per year. Moreover, this energy consumption  is explosively increasing due to the unprecedented expansion of wireless networks to support ubiquitous coverage and connectivity. Furthermore, because of the rapid proliferation of Internet of Things (IoT) applications, most  battery driven power limited IoT devices become useless if their battery power is depleted.  Thus it is critical to use energy in an efficient way or  to harness renewable energy sources.  As remedy, energy harvesting (EH) exploits the pervasive frequency radio signals for replenishing the batteries \cite{X. Lu}.  There have been two research thrusts on EH using RF technology. One focuses on wirelessly powered networks, where a so-called harvest-then-transmit protocol is applied \cite{E. Boshkovska2}. The other one  uses  simultaneous wireless information and power transfer (SWIPT) \cite{Z. Chu}-\cite{D. W. K. Ng1}, which is the focus of this paper. The contributions of SWIPT in CR has been extensively studied. Specifically,  authors of \cite{PVTuan} considered the optimal beamforming design in a MISO CR downlink network. A similar power splitting structure to that of our work is applied at the user side. Hu \emph{et. al} \cite{ZHu}, on the other hand, investigated the objective function of EH energy maximization, and a resource allocation problem was formulated to address that goal. Additionally, \cite{LMojazi} considered the underlay scheme in CR network and proposed the optimal beamforming design.  To address both the SE and EE, a multiple-input single-output (MISO) NOMA CR using SWIPT is considered based on a practical non-linear EH model. Robust beamforming design problems are studied under a pair of channel state information (CSI) error models. The related contributions and the motivation of our work are summarized as follows.

\subsection{Related Work and Motivation}
The prior contributions related to this paper can be divided into two categories based on the  EH model adopted,  i.e. the  linear \cite{D. W. K. Ng1}-\cite{Y. Yuan} and the non-linear EH model \cite{E. Boshkovska2}, \cite{E. Boshkovska}-\cite{Y. Wang}. In  the linear EH model, the power harvested  increases linearly with the input power, while the EH under the non-linear model exhibits more realistic non-linear characteristics especially at the power-tail.

\textbf{Linear EH model:} In \cite{Y. Liu1}, Liu \emph{et al}. analyzed the performance of a cooperative NOMA system relying on SWIPT, which outperformed OMA. Do \emph{et al}. \cite{N. T. Do} extended \cite{Y. Liu1} and studied the beneficial effect of the user selection scheme on the performance of a cooperative NOMA system using SWIPT. In \cite{Z. Yang}, Yang \emph{et al}. presented a theoretical analysis of two power allocation schemes conceived for a cooperative NOMA system with SWIPT. It was shown that the outage probability achieved under NOMA is lower than that obtained under OMA. Diamantoulakis \emph{et al}. \cite{P. D. D3} studied the optimal resource allocation design of wireless-powered NOMA systems. The optimal power and time allocation were designed for maximizing the max-min fairness among users. In their following work \cite{P. D. D4}, a joint downlink and uplink scheme was considered in a wireless powered network, followed by comparisons  between NOMA and TDMA. The results show that NOMA is more energy efficient in the downlink of SWIPT networks.  In order to improve the EE, multiple antennas were applied in a NOMA system associated with SWIPT, and the transmit beamforming and the power splitting factor were jointly optimized for maximizing the transmit rate of users \cite{Y. Xu}.

The contributions in \cite{Y. Liu1}-\cite{Y. Xu} investigated conventional wireless NOMA systems, which did not consider the interference between the secondary network and the primary network. Recently, authors of \cite{D. W. K. Ng1}, \cite{L. Mohjazi}-\cite{F. Zhou3} studied optimal resource allocation problems in CR associated with SWIPT. In \cite{D. W. K. Ng1}, an optimal transmit beamforming scheme was proposed in a multi-objective optimization framework. It was shown that there are several tradeoffs in CR-aided SWIPT. Based on the work in \cite{D. W. K. Ng1}, the authors proposed a jointly optimal beamforming and power splitting scheme to minimize the transmit power of the base station in  multiple-user CR-aided SWIPT \cite{L. Mohjazi}.
Considering the practical imperfect CSI, Zhou \emph{et al}. \cite{F. Zhou2} studied robust beamforming design problems in MISO CR-aided SWIPT, where the bounded and the gaussian CSI error models were applied. It was shown that the performance achieved under the gaussian CSI error model is better than that obtained under the bounded CSI error model. The work in \cite{F. Zhou2} was then extended to multiple-input multiple-output (MIMO) CR-aided SWIPT in \cite{B. Fang} and \cite{C. Xu}, where the bounded CSI error model was applied in \cite{B. Fang} and the gaussian CSI error model was used in \cite{C. Xu} and \cite{replace}. In contrast to \cite{D. W. K. Ng1}, \cite{L. Mohjazi}-\cite{C. Xu}, Zhou \emph{et al}. \cite{F. Zhou3} studied robust resource allocation problems in CR-aided SWIPT under opportunistic spectrum access.

\textbf{Non-linear EH model:} In \cite{E. Boshkovska2}, robust resource allocation schemes were proposed for maximizing the sum transmission rate or the max-min transmission rate of MIMO-assisted wireless powered communication networks, where a practical non-linear EH model is considered. It was shown that a performance gain can be obtained under a practical non-linear EH model over that attained under the linear EH model. In order to maximize the power-efficient and sum-energy harvested by SWIPT systems, Boshkovska \emph{et al}. designed optimal beamforming schemes in \cite{E. Boshkovska} and \cite{E. Boshkovska1}. Recently, under the idealized perfect CSI assumption, the rate-energy region was quantified in MIMO systems relying on SWIPT and the practical non-linear EH model in \cite{K. Xiong}. In order to improve the security of a SWIPT system, a robust beamforming design problem was studied under a bounded CSI error model in \cite{E. Boshkovska3}. The investigations in \cite{E. Boshkovska2}, \cite{E. Boshkovska}-\cite{E. Boshkovska3} were performed in the context of conventional SWIPT systems. Recently, Wang \emph{et al}. \cite{Y. Wang} extended a range of classic resource allocation problems into a wireless powered CR counterpart. The optimal channel and power allocation scheme were proposed for maximizing the sum transmission rate.

The resource allocation schemes proposed in \cite{Y. Liu1}-\cite{Y. Xu} investigated a  conventional NOMA system with SWIPT.  The mutual interference should be considered and the quality of service (QoS) of the PUs should be protected in  NOMA CR. Moreover, the resource allocation schemes proposed in \cite{D. W. K. Ng1}, \cite{L. Mohjazi}-\cite{F. Zhou3} are based on the classic OMA scheme. Thus, these schemes are not applicable to NOMA CR with SWIPT due to the difference between OMA and NOMA. Furthermore, an idealized linear EH model was applied in \cite{D. W. K. Ng1}-\cite{F. Zhou3}, which is impractical since the practical power conversion circuit results in a non-linear end-to-end wireless power transfer. Therefore, it is of great importance to design optimal resource allocation schemes for NOMA CR-aided  SWIPT based on the practical non-linear EH model.

Although the practical non-linear EH model was applied in \cite{E. Boshkovska2}, \cite{E. Boshkovska}-\cite{Y. Wang}, the authors of \cite{E. Boshkovska2}, \cite{E. Boshkovska}-\cite{E. Boshkovska3} considered  conventional OMA systems using SWIPT.  Moreover, the resource allocation scheme proposed in \cite{F. Zhou3} is based on OMA and cannot be directly introduced in NOMA CR-aided SWIPT.  However, at the time of writing, there is a scarcity of investigations on robust resource allocation design for NOMA CR-aided SWIPT under the practical non-linear EH model. Several challenges have to be addressed to design robust resource allocation schemes for NOMA CR-aided SWIPT. For example, the impact of the CSI error and of the residual interference  due to the imperfect SIC should be considered, which makes the robust resource allocation problem quite challenging. Thus, we study robust resource allocation problems in NOMA CR-aided SWIPT.
\subsection{Contributions of the Paper}
Our contribution expands  \cite{E. Boshkovska2} in three major contexts. Firstly, in this paper, a NOMA MISO CR-aided SWIPT is considered, while a OMA MIMO wireless powered network  was used in \cite{E. Boshkovska2}. Secondly, the work in \cite{E. Boshkovska2} relies on the bounded CSI error model, while both the bounded and the gaussian CSI error model are applied in our work. Thirdly, part of our work  considers the minimum transmit power as the optimization objective,  which is not considered in \cite{E. Boshkovska2}. Notice that this paper is also an extension from our conference one \cite{conference} which only considered minimizing transmission power under bounded imperfect CSI model.  The contributions of our work are hence summarized as follows.

\begin{enumerate}
  \item A minimum transmission power problem is formulated under both the bounded and the gaussian CSI error models in a NOMA MISO CR network. The robust beamforming weights and the power splitting ratio are jointly designed. The original problem is hard
to solve owing to its non-convex nature arising from the non-linear EH model as well as owing to the imperfect CSI.  Hence we transform this problem to a convex one. Finally, we prove that the robust beamforming weights can be found and the rank is lower than two under the bounded CSI error model.
  \item We also consider another optimization problem, where the objective function is based on maximizing the harvested energy. Similarly, this problem is formulated under the above pair of imperfect CSI error models. The non-linear EH model makes the original problem even harder to solve. Nevertheless, we managed to transform it to an equivalent form and applied a two-loop procedure for solving it. The inner loop solves a convex problem, while the outer loop iteratively adjusts the parameters. Furthermore, to decouple the coupled variables, a one-dimensional search algorithm is proposed as well.
  \item Simulation results show the superiority of the proposed scheme over the traditional OMA scheme; the performance gain of NOMA becomes higher when the required data rate at each SU is higher. Moreover, the results also demonstrate that under gaussian CSI error model, the performance is generally better than that under the bounded CSI error model.
\end{enumerate}
The remainder of this paper is organized as follows. The system model is presented in Section II. Section III details our robust beamforming design in the context of our power minimization problems under a pair of imperfect CSI error models. Robust beamforming design in EH maximization problems under both imperfect CSI error models are presented in Section IV. Our simulation results are discussed in Section V. Finally, Section VI concludes the paper.

\emph{Notations:} Boldface capital letters and boldface lower case letters denote matrices and vectors, respectively. The identity matrix is denoted by $\mathbf{I}$; vec(\textbf{A}) represents the vectorization of matrix \textbf{A} and it is attained by stacking its column vectors. The Hermitian (conjugate) transpose, trace, and rank of a matrix \textbf{A} are represented respectively by $\mathbf{A^H}$, Tr$\left(\mathbf{A}\right)$ and Rank$\left(\mathbf{A}\right)$. $\mathbf{x}^\dag$ denotes the conjugate transpose of a vector $\mathbf{x}$. $\mathbb{C}^{M\times N}$ denotes a $M$-by-$N$ dimensional complex matrix set. $\mathbf{A}\succeq \mathbf{0} \left(\mathbf{A}\succ \mathbf{0}\right)$ represents that $\mathbf{A}$ is a Hermitian positive semi-definite (definite) matrix.  ${\left\|  \cdot  \right\|}$ represents the Euclidean norm of a vector. ${\left| \cdot \right|}$ denotes the absolute value of a complex scalar. $\mathbf{x} \sim {\cal C}{\cal N}\left( {\mathbf{u},\mathbf{\Sigma } }\right)$ represents that $\mathbf{x}$ is a random vector, which follows a complex Gaussian distribution with mean $\mathbf{u}$ and covariance matrix $\mathbf{\Sigma }$. $\mathbb{E}[ \cdot ]$ represents the expectation operator. ${\rm Re}\left(\mathbf{a}\right)$ extracts the real part of vector $\mathbf{a}$. $\mathbb{R}_{+}$ represents the set of all non-negative real numbers.

\section{System and Energy Harvesting Models}
\subsection{System Model}
We consider a downlink CR system with one cognitive base station (CBS), one primary base station (PBS), $N$ PUs and $K$ SUs. The CBS is equipped with $M$ antennas, while each user and PBS have a single antenna. It is assumed that the SUs are energy-constrained and energy harvest circuits are used. Specifically, the receiver architecture relies on a power splitting design. Once the signal is detected by the receiver, it will be divided into two parts. One part is used for information detection, while the other part for energy harvesting. Similar structures can be found in \cite{Y. Liu1}, \cite{Y. Xu}. To better utilize the radio resources, all UEs are allowed to access the same resource simultaneously. To be specific, the PBS sends messages to all PUs, while the CBS communicates with all SUs simultaneously by applying NOMA principles by controlling the  interference from the CBS to PUs below a certain level \cite{Y. Liu}.
\begin{figure}[h]
\centering
\includegraphics[width=5in]{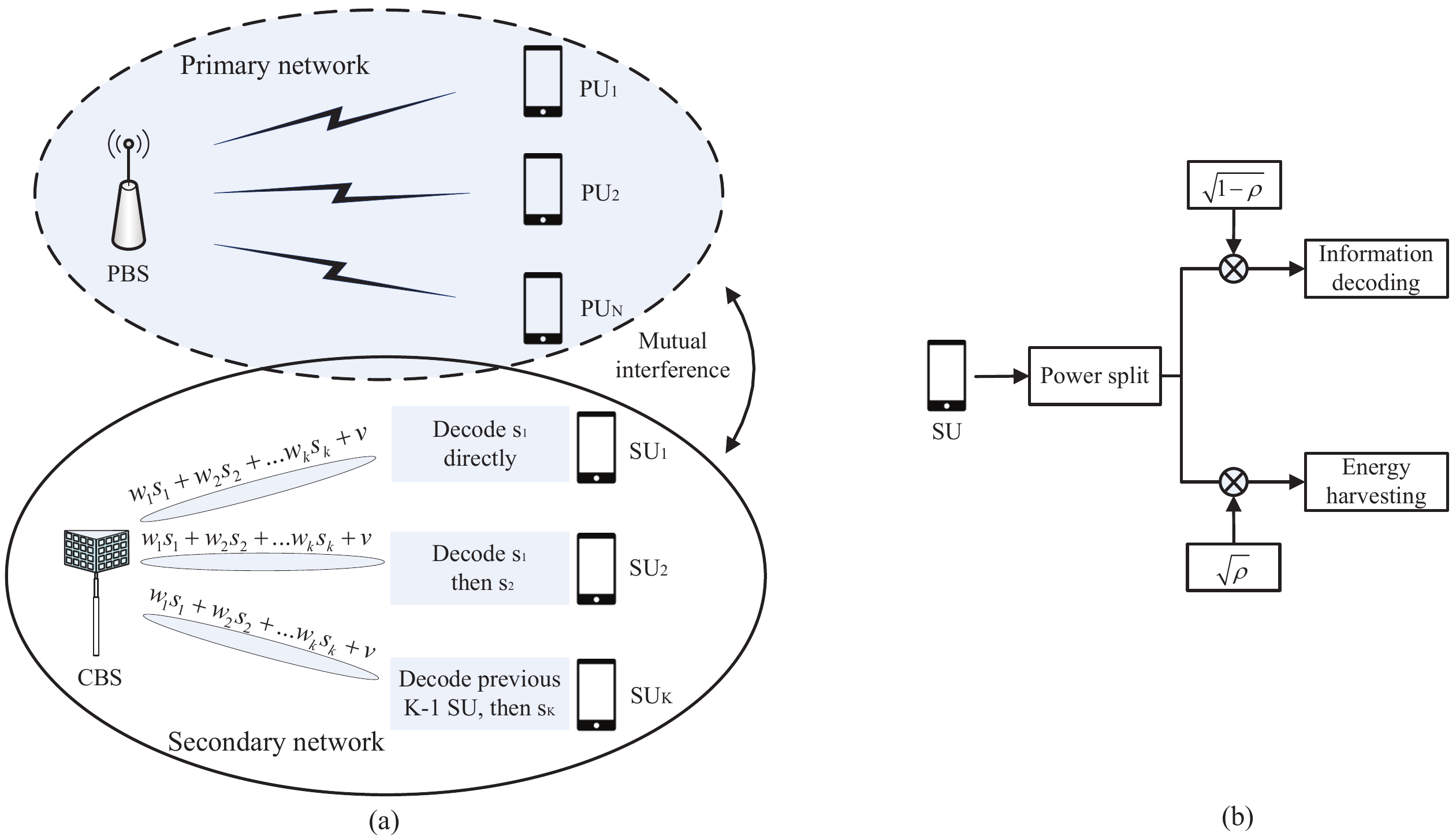}
\caption{(a) an illustration of the system model. (b) the power splitting architecture of SUs.}
\label{model}
\end{figure}
Let us denote the set of SUs and PUs as $\mathcal{K} = \{1,2,\ldots, K\}$ and $\mathcal{N} = \{1,2,\ldots, N\}$, respectively. The signal received by the $k$th SU can be expressed as
\be
y_k^S = \mathbf{h}_k^\dagger \mathbf{x} + n_k^S,\ k \in \mathcal{K},
\ee
where $\mathbf{h}_k \in \mathbb{C}^{M \times 1}$ is the channel gain between the CBS and the $k$th SU, while $n_k^S$ is the joint effect of additive white Gaussian noise (AWGN) and interference from the PBS. $n_k^S \sim \mathcal{CN} (0, \sigma_{k,S}^2)$, where $\sigma_{k,S}^2$ is the power. This interference model represents  a worst-case scenario  \cite{D. W. K. Ng1}. Furthermore, $\mathbf{x}$ is the message transmitted to SUs after precoding. According to the NOMA principle, we have:
\be
\mathbf{x} = \sum_{k=1}^K \mathbf{w}_k s_k + \mathbf{v},
\ee
where $\mathbf{w}_k \in \mathbb{C}^{M \times 1}$ is the precoding vector for the $k$-th UE and $s_k$ is the corresponding intended message. Furthermore, $\mathbf{v} \in \mathbb{C}^{M \times 1}$ is the energy vector allowing us to improve the energy harvesting efficiency at the SUs. We assume that $s_k$ is unitary, i.e. $\mathbb{E} [|s_k|^2] =1$,  and $\mathbf{v}$ obeys the complex Gaussian distribution, i.e.  $\mathbf{v} \sim \mathcal{CN} (0, \mathbf{V})$,  where $\mathbf{V}$ is the covariance matrix of $\mathbf{v}$.

Likewise, the extra interference arriving from the CBS to the $n$-th PU is
\be
y_n^P = \mathbf{g}_n^\dagger \mathbf{x}, \ n \in \mathcal{N},
\ee
where $\mathbf{g}_n^\dagger \in \mathbb{C}^{M \times 1}$ is the channel gain between the CBS and the $n$-th PU \cite{F. Zhou2}.
\subsection{Non-linear EH Model}
Most of the existing literature considered an idealized linear energy harvesting model, where the energy collected by the $k$-th SU is  expressed as $E_k^{\text{Linear}} = \eta  \  E_k^{\text{In}}$, $  E_k^{\text{In}} = \rho  \  \big(\mathbf{h}_k^\dagger (\sum_{j=1}^K \mathbf{w}_j \mathbf{w}_j^\dagger + \mathbf{V}) \mathbf{h}_k + \sigma_{k,S}^2\big)$ is the input power, where $\rho$ is the power splitting factor that controls the amount of received energy allocated to energy harvesting, $0 < \rho < 1$, while $\eta$ is the energy conversion efficiency factor, $ 0 < \eta \leq 1$. However, measurements relying on real-world testbeds show that a typical energy harvesting model exhibits a non-linear end-to-end characteristic. To be specific, the harvested energy first grows almost linearly with the increase of the input power,  and then saturates when the input power reaches a certain level. Several models have been proposed in the literature and one of the most popular ones is \cite{E. Boshkovska2}, which is  formulated as follows:
\begin{subequations}
\begin{eqnarray}
E_k^{\text{Practical}} & = &\frac{\Psi_k^{\text{Practical}} - M_k \Omega_k}{ 1 - \Omega_k}, \ \ \Omega_k = \frac{1}{1+\exp(a_k b_k)},\\
\Psi_k^{\text{Practical}} & = &\frac{M_k}{1+\exp\big(-a_k(E_k^{\text{In}} - b_k)\big)},
\end{eqnarray}
\end{subequations}
where $E_k^{\text{Practical}}$ is the actual energy harvested from the circuit. Furthermore, $\Psi_k^{\text{Practical}}$ represents a  function of the input power $E_k^{\text{In}}$. Additionally, $M_k$ is the maximum power that a receiver can harvest, while  $a_k$ together with $b_k$ characterizes the physical hardware in terms of its circuit sensitivity, limitations, and leakage currents \cite{E. Boshkovska2}.

On the other hand, the signal received in the $k$-th SU information decoding circuit is
\be
y_{k}^D = \sqrt{1-\rho} (\mathbf{h}_k^\dagger \mathbf{x} + n_k^S) + n_{k}^D,
\ee
where $n_{k}^D$ is the AWGN imposed by the information decoding receiver.

\section{Power Minimization Based Problem Formulation}

Since $\mathbf{x}$ is a composite  signal consisting of all SUs' messages, SIC is applied at the receiver side to detect the received signal. The detection is carried out in the same order of the channel gains, i.e. the SUs with lower channel gain will be decoded first. A pair of  imperfect CSI error models are considered, namely a bounded and a gaussian model. We adopt both of these in this paper and assume that all SUs have a perfect knowledge of their own CSI.
\subsection{Bounded CSI Error Model}
In this model, we consider a bounded error imposed on the estimated CSI, which can be treated as the worst-case scenario. Specifically, the channels can be modeled as follows.
\begin{subequations} \label{model1}
\begin{eqnarray}
\mathbf{h}_k &=& \widehat{\mathbf{h}}_k + \Delta  \mathbf{h}_k,  \
\mathbf{\Gamma}_k   \triangleq \big\{   \Delta  \mathbf{h}_k \in \mathbb{C}^{M \times 1} : ||\Delta  \mathbf{h}_k||^2 \leq  \varphi_k^2   \big\}, \ \forall k \in \mathcal{K}, \\
\mathbf{g}_n &=& \widehat{\mathbf{g}}_n + \Delta  \mathbf{g}_n,  \
\mathbf{\Theta}_n   \triangleq  \big\{   \Delta  \mathbf{g}_n \in \mathbb{C}^{M \times 1} : ||\Delta  \mathbf{g}_n||^2 \leq  \psi_n^2   \big\}, \ \forall n \in \mathcal{N},
\end{eqnarray}
\end{subequations}
where $\widehat{\mathbf{h}}_k $ and $\widehat{\mathbf{g}}_n$ are the estimated channel vectors for  $\mathbf{h}_k $ and $\mathbf{g}_n$, respectively, while $\mathbf{\Gamma}_k$ and $\mathbf{\Theta}_n$ define the set of channel variations due to estimation errors. The model defines all the uncertainty regions that are confined by power constraints. Furthermore,  we use block Rayleigh fading channels, which  remain constant within each block, but change from block to block independently.

\subsubsection{NOMA Transmission}

Without loss of generality, we sort the estimated channel of SUs in the ascending order, i.e., $||\hat{\mathbf{h}}_1||^2 \leq ||\hat{\mathbf{h}}_2||^2 \leq \ldots \ ||\hat{\mathbf{h}}_K||^2 $. According to the SIC principle, SU $i$ can detect and remove SU $k$'s signal, for $1 \leq k < i \leq K$. Thus, when SU $i$ decodes signal $s_k$, the signals of the  previous $(k-1) $ SUs have already been removed from the composite received signal. Due to channel estimation errors, however, these $(k-1$) signals may not be completely removed,  leaving some residual signals as interference. Therefore, the signal at UE $i$ when decoding $s_k$ becomes
\begin{equation}
y_{i,k}^S =\sqrt{1-\rho} \big(\mathbf{h}_i^\dagger \mathbf{w}_k s_k + \sum_{j=1}^{k-1} \Delta \mathbf{h}_i^\dagger \mathbf{w}_j s_j + \sum_{j = k+1}^K \mathbf{h}_i^\dagger \mathbf{w}_j s_j +  \mathbf{h}_i^\dagger \mathbf{v} + n_k^S \big) + n_k^D.
\end{equation}
Here, the first term is the desired received signal, the second term is the interference   due to imperfect channel estimation, and the third term represents the NOMA  interference.
The corresponding signal-to-interference-plus-noise ratio (SINR) for the $i$-th SU  after the SIC applied at the receiver is  given by:
\begin{equation}
\text{SINR}_i^k = \frac{ \mathbf{h}_i^\dagger \mathbf{w}_k \mathbf{w}_k^\dagger \mathbf{h}_i }{\sum_{j=1}^{k-1} \Delta \mathbf{h}_i^\dagger \mathbf{w}_j \mathbf{w}_j^\dagger \Delta \mathbf{h}_i + \sum_{j = k+1}^{K}  \mathbf{h}_i^\dagger \mathbf{w}_j \mathbf{w}_j^\dagger \mathbf{h}_i + \mathbf{h}_i^\dagger \mathbf{V} \mathbf{h}_i+ \sigma_{k,S}^2 + \frac{\sigma_D^2}{(1-\rho)}}.
\end{equation}
Since the signal $s_k$ can be detected at every SU $i$, as long as $k < i$, there will be a set of SINRs for signal $s_k$. For CBS, the maximum data rate for SU $k$ should be $ R_k = \log_2 (1+ \min_{k \leq i \leq K} \text{SINR}_i^k ).$
Moreover, the channel estimation error should be considered. The worst-case data rate for SU $k$ becomes
\begin{equation}
R_k = \log_2 \big(1 + \min_{\Delta \mathbf{h}_i \in \mathbf{\Gamma}_i} \ \{ \min_{k \leq i \leq K} \text{SINR}_i^k \}\big).
\end{equation}
 
\subsubsection{Problem Formulation}
In this sub-section, we seek to find the precoding vectors $\mathbf{w}_k,\ k \in \mathcal{K}$, the energy vector $\mathbf{v}$, and the power split ratio $\rho$,  which altogether  achieve a satisfactory  quality of service (QoS) for all users, and  at the same time,  they can harvest part of the energy for their future usage. Thus, the problem can be formulated as follows:
\begin{subequations}
\begin{eqnarray}
 \mathbf{P}_1 :  & &  \min_{\mathbf{w}_k  \in \mathbb{C}^{M \times 1}, \mathbf{V} \in \mathbb{C}^{M \times M}, \rho}  \   \text{Tr} (\sum_{k=1}^K \mathbf{w}_k^\dagger \mathbf{w}_k) + \text{Tr}(\mathbf{V})  \\
\text{s.t.} \   C1 &:& R_k \geq R_{k,\text{min}},  \ \forall k \in \mathcal{K}, \label{p1c1}\\
C2 &:& E_k^{\text{Practical}} \ \geq P_{k,s},  \ \forall \Delta  \mathbf{h}_k \in  \mathbf{\Gamma}_k, \ \forall k \in \mathcal{K}, \\
C3 &:& \mathbf{g}_n^\dagger \big(  \sum_{j=1}^K \mathbf{w}_j \mathbf{w}_j^\dagger + \mathbf{V} \big) \mathbf{g}_n \leq P_{n,p}, \ \forall \Delta  \mathbf{g}_n \in  \mathbf{\Theta}_n, \ \forall n \in \mathcal{N}, \\
C4 &:& \sum_{j=1}^K \mathbf{w}_j^\dagger \mathbf{w}_j + \text{Tr} (\mathbf{V}) \leq P_B, \\
C5 &:& 0 < \rho < 1, \\
C6 &:& \mathbf{V} 	\succeq \mathbf{0}.
\end{eqnarray}
\end{subequations}

Our goal is to minimize the total transmitted power. The constraint $C1$ ensures that  SU $k$ does attain the predefined minimum data rate; $C2$  allows each SU to harvest the amount of energy that at least compensates the static power dissipation $P_{k,s}$; $C3$ is the interference limit for the $n$-th PU; $C4$ represents the  maximum transmit power constraint of the BS;  in $C5$, the power split factor should be in the range of $(0,1)$. The  optimization problem $\mathbf{P}_1$ is hard to solve due to its non-convexity constraints $C1$ and $C2$. Moreover, the realistic imperfect CSI imposes another challenge on the original problem. In the following, we transform the variables.

Let us introduce $\mathbf{W}_k = \mathbf{w}_k \mathbf{w}_k^\dagger$ and $\gamma_{k,\text{min}} \triangleq (2^ { R_{k,\text{min}}}-1) $. Then $C1$ in (\ref{p1c1}) becomes
\bea
\label{C1}
 \min_{\Delta \mathbf{h}_i \in \mathbf{\Gamma}_i} \  \frac{ \mathbf{h}_i^\dagger \mathbf{W}_k  \mathbf{h}_i  }{\sum_{j=1}^{k-1} \Delta \mathbf{h}_i^\dagger \mathbf{W}_j \Delta \mathbf{h}_i + \sum_{j = k+1}^{K} \mathbf{h}_i^\dagger \mathbf{W}_j \mathbf{h}_i  + \mathbf{h}_i^\dagger \mathbf{V} \mathbf{h}_i+  \sigma_{k,S}^2 + \frac{\sigma_D^2}{(1-\rho)}} \geq \gamma_{k,\text{min}}, \\ \nonumber \ i = \{k, k+1, \ldots, K \}, \forall k \in \mathcal{K}.
\eea
For the notational simplicity, we denote the above constraint as $\Xi_{i,k}$. Thus, $\mathbf{P}_1$ becomes
\begin{subequations}
\begin{eqnarray}
 \mathbf{P}_2 :  & &  \min_{\mathbf{W}_k  \in \mathbb{C}^{M \times M}, \mathbf{V} \in \mathbb{C}^{M \times M}, \rho}  \   \text{Tr}(\sum_{k=1}^K \mathbf{W}_k  + \mathbf{V})  \\
\text{s.t.} \   C1 &:& \Xi_{i,k}  \label{p2c1}\\
C2 &:& E_k^{\text{Practical}} \ \geq P_{k,s}, \ \forall \Delta  \mathbf{h}_k \in  \mathbf{\Gamma}_k, \ \forall k \in \mathcal{K},\label{p2c2}\\
C3 &:& \mathbf{g}_n^\dagger \big(  \sum_{j=1}^K \mathbf{W}_j + \mathbf{V} \big) \mathbf{g}_n \leq P_{n,p}, \ \forall \Delta  \mathbf{g}_n \in  \mathbf{\Theta}_n, \ \forall n \in \mathcal{N}, \label{p2c3}\\
C4 &:& \text{Tr}(\sum_{k=1}^K \mathbf{W}_k  + \mathbf{V}) \leq P_B, \label{p2c4}\\
C5 &:& 0 < \rho < 1, \label{p2c5}\\
C6 &:& \mathbf{V} 	\succeq \mathbf{0}, \mathbf{W}_k  \succeq \mathbf{0}, \label{p2c6}\\
C7 &:& \text{Rank} (\mathbf{W}_k) = 1, \ \forall k \in \mathcal{K} \label{p2c7}.
\end{eqnarray}
\end{subequations}
Here, $C6$ comes from the fact that both $\mathbf{V}$ and $\mathbf{W}_k$ are positive semi-definite matrices. The extra constraint that  the rank of $\mathbf{W}_k$ should be 1 is also non-convex. In what follows, we first reformulate  $C1$ in (\ref{p2c1}) according to the  $\mathcal{S}$-Procedure of \cite{convex}.
\begin{lemma}
$C1$ in (\ref{p2c1}) can be reformulated as
\bea
\begin{bmatrix}
  {\begin{array}{cc}
   \alpha_{i,k} \mathbf{I} + \mathbf{C}_k - \gamma_{k,\text{min}} \sum_{j=1}^{k-1} \mathbf{W}_j & \mathbf{C}_k \widehat{\mathbf{h}_i} \\
   \widehat{\mathbf{h}_i}^\dagger \mathbf{C}_k  & -\alpha_{i,k} \varphi_k^2 + \mathbf{\Phi}_k \\
  \end{array} }
  \end{bmatrix}   \succeq  \mathbf{0}  ,\ \forall k \in \mathcal{K}, \ i = \{k, k+1, \ldots, K\} \label{newC1},
\eea
where $\mathbf{C}_k = \mathbf{W}_k -\gamma_{k,\text{min}} (\sum_{j=k+1}^K \mathbf{W}_j +  \mathbf{V})$ and $\mathbf{\Phi}_k =  \widehat{\mathbf{h}_i}^\dagger \mathbf{C}_k \widehat{\mathbf{h}_i} - \gamma_{k,\text{min}} \big( \sigma_{k,S}^2 + \frac{\sigma_D^2}{(1-\rho)}\big)$, and $\alpha_{i,k}$ is a slack variable conditioned on $\alpha_{i,k} \geq 0$.
\end{lemma}

\begin{IEEEproof}
Given $\mathbf{h}_i = \widehat{\mathbf{h}_i} + \Delta \mathbf{h}_i $ and  (\ref{C1}), we have
\bea
\Delta \mathbf{h}_i^\dagger \big( \gamma_{k,\text{min}}( \sum_{j \neq k}\mathbf{W}_j + \mathbf{V}) - \mathbf{W}_k  \big) \Delta \mathbf{h}_i  + 2 \ \text{Re} \big \{ \widehat{\mathbf{h}_i}^\dagger \big( \gamma_{k,\text{min}} (\sum_{j=k+1}^{K} \mathbf{W}_j + \mathbf{V}) - \mathbf{W}_k \big)\Delta \mathbf{h}_i \big \} \\ \nonumber
+ \widehat{ \mathbf{h}_i}^\dagger  \big( \gamma_{k,\text{min}} (\sum_{j=k+1}^{K} \mathbf{W}_j + \mathbf{V}) - \mathbf{W}_k \big) \widehat{ \mathbf{h}_i} + \gamma_{k,\text{min}} \big( \sigma_{k,S}^2 + \frac{\sigma_D^2}{(1-\rho)} \big) \leq 0.
\eea
From the fact that $\Delta \mathbf{h}_i^\dagger \Delta \mathbf{h}_i - \varphi_k^2 \leq 0 $ and according to the $\mathcal{S}$-Procedure, the lemma is proved.
\end{IEEEproof}

Similarly, $C3$ in (\ref{p2c3}) can be transformed into
\bea
\begin{bmatrix}
 {\begin{array}{cc}
   \beta_{n} \mathbf{I} - \mathbf{\Sigma} & -\mathbf{\Sigma} \widehat{\mathbf{g}}_n \\
   -\widehat{\mathbf{g}}_n^\dagger \mathbf{\Sigma}  & -\beta_{n} \psi_n^2 - \widehat{\mathbf{g}}_n^\dagger \mathbf{\Sigma} \widehat{\mathbf{g}}_n + P_{n,p} \\
  \end{array} }
 \end{bmatrix}  \succeq  \mathbf{0},  \ \forall n \in \mathcal{N}  \label{newC3},
\eea
where $\mathbf{\Sigma} = \sum_{j=1}^K \mathbf{W}_j + \mathbf{V}$, and $\beta_{n} \geq 0$ is also a slack variable.

Next, we apply similar manipulations to (\ref{p2c2}), which becomes
\bea
\min_{\Delta \mathbf{h}_k \in \mathbf{\Gamma}_k} \rho  \ \big( \mathbf{h}_k^\dagger \mathbf{\Sigma}  \mathbf{h}_k + \sigma_{k,S}^2\big) &\geq& D_k , \label{nonlinear}
\eea
where $D_k = -  \ln \big( \frac{1}{P_{k,s}  (1-\Omega_k)/M_k + \Omega_k}-1\big) /a_k + b_k$ is a constant. This condition holds, provided that $a_k > 0$,  which is always true in real systems.

Then,  applying  the $\mathcal{S}$-Procedure to (\ref{nonlinear}), we have  the following
\bea
\begin{bmatrix}
 {\begin{array}{cc}
   \theta_{k} \mathbf{I} + \mathbf{\Sigma}  & \mathbf{\Sigma} \widehat{\mathbf{h}}_k \\
   \widehat{\mathbf{h}}_k^\dagger \mathbf{\Sigma}  & -\theta_{k} \varphi_k^2 + \widehat{\mathbf{h}}_k^\dagger \mathbf{\Sigma}  \widehat{\mathbf{h}}_k + \sigma_{k,S}^2 - \frac{D_k}{\rho} \\
  \end{array} }
 \end{bmatrix}  \succeq  \mathbf{0},  \ \forall k \in \mathcal{K}  \label{newC2},
\eea
where $\theta_k \geq 0$.

Therefore, $\mathbf{P}_2$ becomes
\begin{subequations}
\begin{eqnarray}
 \mathbf{P}_3 :  & &  \min_{\mathbf{W}_k , \mathbf{V}, \rho, \{\alpha_{i,k}\}, \{\beta_{n}\}, \{\theta_{k}\}}  \   \text{Tr}(\sum_{k=1}^K \mathbf{W}_k  + \mathbf{V})  \\
\text{s.t.} & & \   (\ref{newC1}), (\ref{newC3}), (\ref{newC2}), (\ref{p2c4}), (\ref{p2c5}), (\ref{p2c6}),  \\
 & & \alpha_{i,k}, \beta_{n}, \theta_{k} \geq 0 ,\ \forall k \in \mathcal{K}, \ i = \{k,k+1,\ldots, K\}, \ \forall n \in \mathcal{K}.
\end{eqnarray}
\end{subequations}
Observe that we drop (\ref{p2c7}), since it is not a convex term. This relaxation is commonly referred to as the semi-definite relaxation (SDR) technique. For the specific problem in $\mathbf{P}_2$, the following theorem proves that the optimal $\mathbf{W}_k$ has a limited rank.
\begin{theorem}
If $\mathbf{P}_2$ is feasible, the rank of $\mathbf{W}_k, k \in \mathcal{K}$ is always less than or equal to 2.
\end{theorem}

\begin{IEEEproof}
See Appendix.
\end{IEEEproof}

The transformed problem $\mathbf{P}_3$ is not convex  because of the coupling variables $\rho$ in (\ref{newC2}) and $(1-\rho)$ in the denominator of (\ref{newC1}). To be able to take advantage of the \emph{CVX} software package, we introduce a pair of auxiliary variables. Specifically, let $p = \frac{1}{1-\rho}$ and $q = \frac{1}{\rho}$. In this way, (\ref{newC1}), (\ref{newC3}), and (\ref{newC2}) become convex terms. Then, we have additional constraints for $p$ and $q$:
\bea \label{newTrans}
p \geq \frac{1}{1-\rho} \ \  \text{and} \ \  q \geq \frac{1}{\rho}.
\eea
It may be readily verified that this transformation does not change the optimal solution of $\mathbf{P}_3$.

\subsection{Matrix Decomposition}
Now we proceed to  find the solution of the problem $\mathbf{P}_2$, after which there is one more step to get the original solution for $\mathbf{w}_k$. If $\mathbf{W}_k$ yields rank 1, we can simply write $\mathbf{W}_k^\star = \mathbf{w}_k^\star \mathbf{w}_k^{\star \dagger}$. Otherwise, if $\text{Rank}(\mathbf{W}_k^\star) = 2$, we have several optional approaches to extract $\mathbf{w}_k^\star$. To name a few, we list two methodologies here.
\begin{enumerate}
\item \emph{Eigen-decomposition.} Let us denote two eigenvalues of $\mathbf{W}_k^\star$ by $\lambda_1$ and $\lambda_2$, where $\lambda_1 > \lambda_2 \geq 0$. Clearly, $\mathbf{W}_k^\star = \lambda_1 \mathbf{w}_{1k} \mathbf{w}_{1k}^{\dagger} + \lambda_2 \mathbf{w}_{2k} \mathbf{w}_{2k}^{\dagger}$,  $\mathbf{w}_{ik}, i=\{1,2\}$ are the corresponding eigenvectors. To get the rank 1 approximation from a rank 2 matrix,  we can let the solution of the original problem be $\widehat{\mathbf{w}}_k = \sqrt{\lambda_1} \mathbf{w}_{1k} \mathbf{w}_{1k}^{\dagger}$, provided it is feasible.
\item \emph{Randomization technique.} Similar to eigen-decomposition, we first decompose $\mathbf{W}_k^\star$ according to $\mathbf{W}_k^\star = \mathbf{U}_k \mathbf{T}_k \mathbf{U}_k^\dagger$. Then, we let $\widehat{\mathbf{w}}_k =\mathbf{U}_k \mathbf{T}_k^{1/2} \mathbf{e}_k$, where the \emph{m}-th element of $\mathbf{e}_k$ is $[\mathbf{e}_k]_m = e^{j\theta_{k,m}}$ and  $\theta_{k,m}$  obeys an independent and uniform distribution within $[0,2\pi)$.
\end{enumerate}
The above two methods are essentially the same. If we want to get a more precise result, another scaling factor can be added. Specifically, let us define $c_k$ as the scaling factor yet to be determined. Certainly, the problem can be transformed in terms of $\mathbf{W}_k$ and $c_k$, once we get the optimal value, we can apply either one of the above methods to get a better result. Another point worth noting here is that when the rank of $\mathbf{W}_k$ is 2, there only exists the approximation result of $\mathbf{w}_k^{\star}$, and this approximation always provides an upper bound. 
\subsection{Gaussian CSI Error Model}
In Section III-\emph{A}, we introduced a bounded channel model, which defines a confined region for the channel variations, which provides a worst-case estimation.  Another commonly used  more realistic  estimation model assumes that the channel estimation error obeys the  Gaussian distribution \cite{F. Zhou2}\cite{Y. Yuan}\cite{Y. Wang}, which is formulated as follows:
\begin{subequations}
\begin{eqnarray}
\mathbf{h}_k &=& \widehat{\mathbf{h}}_k + \Delta  \mathbf{h}_k,  \ \Delta \mathbf{h}_k \sim \mathcal{CN}(0, \mathbf{H}_k), \ \forall k \in \mathcal{K}, \\
\mathbf{g}_n &=& \widehat{\mathbf{g}}_n + \Delta  \mathbf{g}_n,  \ \Delta \mathbf{g}_n \sim \mathcal{CN}(0, \mathbf{G}_n), \ \forall n \in \mathcal{N},
\end{eqnarray}
\end{subequations}
where $\Delta \mathbf{h}_k $ and $\Delta \mathbf{g}_n$ are the channel estimation error vectors, while $\widehat{\mathbf{h}}_k$ and $\widehat{\mathbf{g}}_n$ are the channel vectors estimated at the BS side. Furthermore, $\mathbf{H}_k$ and $\mathbf{G}_n$ are the covariance matrices of the estimation error vectors.

Even though we apply different channel models, the residual interference  due to imperfect CSI estimation affects the  message detection similarly to the bounded error model. Thus the achievable data rate expression of SU $k$ remains the same except that $\Delta \mathbf{h}_k$ is in a new set. In contrast to the existing NOMA contributions on imperfect CSI \cite{sun1}, in this paper we use the above-mentioned gaussian estimation error model to form an optimization problem as follows:
\begin{subequations}
\begin{eqnarray}
 \mathbf{P}_4 :  & &  \min_{\mathbf{W}_k  \in \mathbb{C}^{M \times M}, \mathbf{V} \in \mathbb{C}^{M \times M}, \rho}  \   \text{Tr}(\sum_{k=1}^K \mathbf{W}_k  + \mathbf{V})  \\
\text{s.t.} \   C1 &:& \text{Pr} \{ R_k \geq R_{k,\text{min}} \} \geq 1- \xi_k ,  \ \forall k \in \mathcal{K},\label{p4c1}\\
C2 &:& \text{Pr} \{ E_k^{\text{Practical}} \ \geq P_{k,s} \} \geq 1- \xi_{k,s},  \ \forall \Delta  \mathbf{h}_k \sim \mathcal{CN}(0, \mathbf{H}_k) , \ \forall k \in \mathcal{K}, \label{p4c2}\\
C3 &:& \text{Pr} \big\{ \mathbf{g}_n^\dagger \mathbf{\Sigma} \mathbf{g}_n \leq P_{n,p} \big\} \geq 1- \xi_{n,p},  \forall \Delta  \mathbf{g}_n \sim \mathcal{CN}(0, \mathbf{G}_n), \forall n \in \mathcal{N}, \label{p4c3}\\
C4 &:& (\ref{p2c4}) - (\ref{p2c7}).
\end{eqnarray}
\end{subequations}

Here, we assume that the probability of having a rate of $R_k$ is higher than $R_{k,\text{min}}$, which is a predefined value, and we use the threshold $\xi_k$ to control the probability. Likewise, $\xi_{k,s}$ and $\xi_{n,p}$, where $k \in \mathcal{K}$ and $n \in \mathcal{N}$,  are used for controlling the outage probability of harvested energy of the $k$th SU and the interference experienced by the $n$-th PU, respectively. $\mathbf{P}_4$ is hard to solve owing to its non-convexity, together with constraints $C1-C3$,  which involve probability and uncertainty. Inspired by \cite{F. Zhou2}, we solve the resulted optimization problem with the aid of approximations by applying Bernstein-type inequalities \cite{prob1}.

\subsubsection{Bernstein-type Inequality I  \cite{prob1}}Let $f(\mathbf{z}) = \mathbf{z}^\dagger \mathbf{A} \mathbf{z} + 2 \text{Re}\{\mathbf{z}^\dagger \mathbf{b}\} + c$, where $\mathbf{A}\in \mathbb{H}^{N}$, $\mathbf{b} \in \mathbb{C}^{N \times 1}$, $c \in \mathbb{R}$, and $\mathbf{z} \sim \mathcal{CN}(0, \mathbf{I})$. For any $\xi \in (0,1]$, an approximate and convex form of
\be
\text{Pr} \{f(\mathbf{z}) \geq 0 \} \geq 1- \xi
\ee
can be written as
\begin{subequations} \label{outage}
\begin{eqnarray}
\text{Tr} (\mathbf{A}) - \sqrt{-2 \ln (\xi)} \upsilon_1 + \ln (\xi) \upsilon_2 + c \geq 0,  \\
\Bigg  | \Bigg |\begin{bmatrix}
 {\begin{array}{c}
   \text{vec}(\mathbf{A})  \\
   \sqrt{2} \mathbf{b}  \\
  \end{array} }
 \end{bmatrix} \Bigg | \Bigg| \leq \upsilon_1, \\
\upsilon_2 \mathbf{I} + \mathbf{A} \succeq \mathbf{0}, \upsilon_2 \geq 0.
\end{eqnarray}
\end{subequations}
Here, $\upsilon_1$ and $\upsilon_2$ are slack variables.

In order to use the above Lemma, we have to transform $\Delta \mathbf{h}_i$ to a standard complex Gaussian vector. Let $\Delta \mathbf{h}_i = \mathbf{H}_i^{1/2} \tilde {\mathbf{h}}_i$, where $\tilde {\mathbf{h}}_i \sim \mathcal{CN}(0, \mathbf{I})$. Substituting it into (\ref{C1}), the convex approximation becomes
\begin{subequations}\label{p4c1_trans}
\begin{eqnarray}
\text{Tr} \big(\mathbf{H}_i^{1/2} (\mathbf{C}_k - \gamma_{k,\text{min}} \sum_{j=1}^{k-1} \mathbf{W}_j) \mathbf{H}_i^{1/2} \big) - \sqrt{-2 \ln (\xi_k)} \upsilon_{1i,k} + \ln (\xi_k) \upsilon_{2i,k} + c_{i,k} \geq 0,  \\
c_{i,k} = \widehat{\mathbf{h}}_i^\dagger \mathbf{C}_k \widehat{\mathbf{h}}_i - r_{k,\text{min}}\big(\sigma_{k,S}^2 + \frac{\sigma_D^2}{1-\rho}\big), \label{prob_C2}\\
\Bigg  | \Bigg |\begin{bmatrix}
 {\begin{array}{c}
   \text{vec}\big(\mathbf{H}_i^{1/2} (\mathbf{C}_k - \gamma_{k,\text{min}} \sum_{j=1}^{k-1} \mathbf{W}_j) \mathbf{H}_i^{1/2} \big)  \\
   \sqrt{2} \mathbf{H}_i^{1/2} \mathbf{C}_k \widehat{\mathbf{h}}_i  \\
  \end{array} }
 \end{bmatrix} \Bigg | \Bigg| \leq \upsilon_{1i,k}, \\
\upsilon_{2i,k}  \mathbf{I} + \big(\mathbf{H}_i^{1/2} (\mathbf{C}_k - \gamma_{k,\text{min}} \sum_{j=1}^{k-1} \mathbf{W}_j) \mathbf{H}_i^{1/2} \big) \succeq \mathbf{0}, \upsilon_{2i,k}  \geq 0, \ \forall k \in \mathcal{K},  i = \{k,\ldots, K\},
\end{eqnarray}
\end{subequations}
where $\upsilon_{1i,k}$ and $\upsilon_{2i,k}$ are slack variables.

For (\ref{p4c2}), we use  a simple transformation similar as that in  (\ref{nonlinear}), which leads to:
\be
\text{Pr} \big\{ \rho   ( \mathbf{h}_k^\dagger \mathbf{\Sigma} \mathbf{h}_k + \sigma_{k,S}^2) \geq  D_k\big \} \geq 1-\xi_{k,s}. \label{nonlinear_inP4}
\ee
Furthermore, by applying the inequalities in (\ref{outage}), (\ref{nonlinear_inP4}) can be expressed as
\begin{subequations}\label{p4c2_trans}
\begin{eqnarray}
\text{Tr} \big(\mathbf{H}_k^{1/2}\mathbf{\Sigma}  \mathbf{H}_k^{1/2} \big) - \sqrt{-2 \ln (\xi_{k,s})} \upsilon_{1k,s} + \ln (\xi_{k,s}) \upsilon_{2k,s} + c_{k,s} \geq 0,  \\
c_{k,s} = \widehat{\mathbf{h}}_k^\dagger \mathbf{\Sigma}  \widehat{\mathbf{h}}_k + \sigma_{k,S}^2 - \frac{D_k}{\rho}, \label{prob_C1} \\
\Bigg  | \Bigg |\begin{bmatrix}
 {\begin{array}{c}
   \text{vec}\big( \mathbf{H}_k^{1/2}\mathbf{\Sigma}  \mathbf{H}_k^{1/2} \big)  \\
   \sqrt{2} \mathbf{H}_k^{1/2}  \mathbf{\Sigma}  \widehat{\mathbf{h}}_k  \\
  \end{array} }
 \end{bmatrix} \Bigg | \Bigg| \leq \upsilon_{1k,s}, \\
\upsilon_{2k,s}  \mathbf{I} + \big(\mathbf{H}_k^{1/2}\mathbf{\Sigma}  \mathbf{H}_k^{1/2} \big) \succeq \mathbf{0}, \upsilon_{2k,s}  \geq 0, \ \forall k \in \mathcal{K},
\end{eqnarray}
\end{subequations}
where $\upsilon_{1k,s}$ and $\upsilon_{2k,s}$, $k \in \mathcal{K}$, are slack variables.

\subsubsection{Bernstein-type Inequality II \cite{prob2}}  Let $f(\mathbf{z}) = \mathbf{z}^\dagger \mathbf{A} \mathbf{z} + 2 \text{Re}\{\mathbf{z}^\dagger \mathbf{b}\} + c$, where $\mathbf{A}\in \mathbb{H}^{N}$, $\mathbf{b} \in \mathbb{C}^{N \times 1}$, $c \in \mathbb{R}$, and $\mathbf{z} \sim \mathcal{CN}(0, \mathbf{I})$. For any $\xi \in (0,1]$, an approximate and convex form for
\be
\text{Pr} \{f(\mathbf{z}) \leq 0 \} \geq 1- \xi
\ee
can be written as
\begin{subequations}
\begin{eqnarray}
\text{Tr} (\mathbf{A}) + \sqrt{-2 \ln (\xi)} \upsilon_1 - \ln (\xi) \upsilon_2 + c \geq 0,  \\
\Bigg  | \Bigg |\begin{bmatrix}
 {\begin{array}{c}
   \text{vec}(\mathbf{A})  \\
   \sqrt{2} \mathbf{b}  \\
  \end{array} }
 \end{bmatrix} \Bigg | \Bigg| \leq \upsilon_1, \\
\upsilon_2 \mathbf{I} - \mathbf{A} \succeq \mathbf{0}, \upsilon_2 \geq 0,
\end{eqnarray}
\end{subequations}
where $\upsilon_1$ and $\upsilon_2$ are slack variables.

We apply Bernstein-type Inequality II to (\ref{p4c3}), and let $\Delta \mathbf{g}_n= \mathbf{G}_n^{1/2} \tilde {\mathbf{g}}_n$, where $\tilde {\mathbf{g}}_n \sim \mathcal{CN}(0, \mathbf{I})$ is a standard Gaussian vector. We can have the following convex-form approximation.
\begin{subequations}\label{p4c3_trans}
\begin{eqnarray}
\text{Tr} (\mathbf{G}_n^{1/2} \mathbf{\Sigma} \mathbf{G}_n^{1/2} ) + \sqrt{-2 \ln (\xi_{n,p})} \upsilon_{1,n} - \ln (\xi_{n,p}) \upsilon_{2,n} + c_n \geq 0,  \\
c_n = \widehat{\mathbf{g}}_n^\dagger \mathbf{\Sigma} \widehat{\mathbf{g}}_n - P_{n,p}, \\
\Bigg  | \Bigg |\begin{bmatrix}
 {\begin{array}{c}
   \text{vec}(\mathbf{G}_n^{1/2} \mathbf{\Sigma} \mathbf{G}_n^{1/2})  \\
   \sqrt{2} \mathbf{G}_n^{1/2} \mathbf{\Sigma} \widehat{\mathbf{g}}_n  \\
  \end{array} }
 \end{bmatrix} \Bigg | \Bigg| \leq \upsilon_{1,n}, \\
\upsilon_{2,n} \mathbf{I} - \mathbf{G}_n^{1/2} \mathbf{\Sigma} \mathbf{G}_n^{1/2} \succeq \mathbf{0}, \upsilon_{2,n} \geq 0, \forall n \in \mathcal{N},
\end{eqnarray}
\end{subequations}
where $\upsilon_{1,n}$ and $\upsilon_{2,n}$ are slack variables.

Lastly, we relax $\mathbf{P}_4$ by dropping the constraint that $\mathbf{W}_k$ should have rank 1 for now, since it is not a convex one. The relaxed version of the problem is
\begin{subequations}
\begin{eqnarray}
 \mathbf{P}_5 :  & &  \min_{\mathbf{W}_k , \mathbf{V}, \rho, \{\upsilon_{1i,k}\},\{\upsilon_{2i,k}\},\{\upsilon_{1k,s}\},\{\upsilon_{2k,s}\}, \{\upsilon_{1,n}\}, \{\upsilon_{2,n}\} }  \   \text{Tr}(\sum_{k=1}^K \mathbf{W}_k  + \mathbf{V})  \\
\text{s.t.} & & \   (\ref{p4c1_trans}), (\ref{p4c2_trans}), (\ref{p4c3_trans}), (\ref{p2c4}), (\ref{p2c5}), (\ref{p2c6}).
\end{eqnarray}
\end{subequations}
Likewise, the coupling variables in (\ref{prob_C2}) and (\ref{prob_C1}) make $\mathbf{P}_5$ a non-convex problem. Thus we can  still use the transformation in (\ref{newTrans}), which converts  $\mathbf{P}_5$ into an equivalent optimization problem  that can be efficiently solved by \emph{CVX}.

\section{Maximum Harvested Energy Problem Formulation}
In contrast to Sections III, where the minimum transmission power problem is considered, in the following we consider the optimization problem of maximizing the total harvested energy. This problem has important real-world applications, since most of the consumer electronics products are battery-driven and thus their energy efficiency is critical. In this section, we first formulate the problem, then we transform it in a convex way so that an existing software package can solve  it efficiently. A one-dimensional search algorithm will be used. Furthermore, we also consider our previous pair of channel models.

\subsection{Bounded CSI Error Model}
Upon considering the imperfect CSI model used in (\ref{model1}), the maximum total harvested energy of all SUs can be formulated as follows:
\begin{subequations}
\begin{eqnarray}
 \mathbf{P}_6 :  & &  \max_{\mathbf{W}_k  \in \mathbb{C}^{M \times M}, \mathbf{V} \in \mathbb{C}^{M \times M}, \rho, \{\alpha_{i,k}\}, \{\beta_{n}\}, \{\theta_{k}\}}  \   \ \sum_{k=1}^{K} \  E_k^{\text{Practical}}  \\
 \text{s.t.} & & \   (\ref{newC1}), (\ref{newC3}), (\ref{p2c4}), (\ref{p2c5}), (\ref{p2c6}), (\ref{p2c7}),   \\
 & & \alpha_{i,k}, \beta_{n}, \theta_{k} \geq 0 ,\ \forall k \in \mathcal{K}, \ i = \{k,k+1,\ldots, K\}, \ \forall n \in \mathcal{K}.
\end{eqnarray}
\end{subequations}
The rank operation is not convex, thus we drop the constraint (\ref{p2c7}) first, as previously in $\mathbf{P}_3$. Additionally, the objective function relies on a realistic non-linear energy harvesting model, and it is not convex either. Essentially, it is a sum-of-ratio problem, and its global optimization is possible by applying the following transformations:
\begin{subequations}
\begin{eqnarray}
& & \max_{\mathbf{W}_k  \in \mathbb{C}^{M \times M}, \mathbf{V} \in \mathbb{C}^{M \times M}, \rho, \{\alpha_{i,k}\}, \{\beta_{n}\}, \{\theta_{k}\}, \{\tau_k\}}  \   \ \sum_{k=1}^{K} \  \frac{M_k}{1+\exp\big(-a_k(\tau_k - b_k)\big)}, \label{p6_obj1}\\
& & E_k^{\text{In}}  \geq \tau_k, \ \forall \Delta  \mathbf{h}_k . \ \forall k \in \mathcal{K} \label{p6_obj2}.
\end{eqnarray}
\end{subequations}
After applying  the $\mathcal{S}$-Procedure of \cite{convex} to (\ref{p6_obj2}), it becomes
\bea
\begin{bmatrix}
 {\begin{array}{cc}
   \theta_{k} \mathbf{I} + \mathbf{\Sigma}  & \mathbf{\Sigma} \widehat{\mathbf{h}}_k \\
   \widehat{\mathbf{h}}_k^\dagger \mathbf{\Sigma}  & -\theta_{k} \varphi_k^2 + \widehat{\mathbf{h}}_k^\dagger \mathbf{\Sigma}  \widehat{\mathbf{h}}_k + \sigma_{k,S}^2 - \frac{\tau_k}{\rho} \\
  \end{array} }
 \end{bmatrix}  \succeq  \mathbf{0},  \ \forall k \in \mathcal{K} \label{p6_newobj}.
\eea
Furthermore, according to \cite{E. Boshkovska}, \cite{Y. Jong}, if $\mathbf{P}_6$ has the optimal solutions $\mathbf{W}_k^{\star}$ and $\mathbf{V}^{\star}$, there exist two sets of vectors $\bm{\mu} = \{\mu_1, \mu_2, \ldots, \mu_K \}$ and $\bm{\epsilon} = \{\epsilon_1, \epsilon_2, \ldots, \epsilon_K \}$ such that the solutions are also optimal for the following equivalent parametric optimization problem:
\be
\mathbf{P}_7 : \max_{\mathbf{W}_k  \in \mathbb{C}^{M \times M}, \mathbf{V} \in \mathbb{C}^{M \times M}, \rho, \{\alpha_{i,k}\}, \{\beta_{n}\}, \{\theta_{k}\}, \{\tau_k\}}  \   \ \sum_{k=1}^{K} \  \mu_k \big\{ M_k - \epsilon_k \big(1+\exp (-a_k(\tau_k - b_k) ) \big) \big\}.
\ee
The optimal solutions and the vectors should satisfy
\begin{subequations}\label{constraint_p6}
\begin{eqnarray}
\epsilon_k \big(1+\exp (-a_k(\tau_k^\star - b_k) ) \big) - M_k &=& 0, \\
\mu_k \big(1+\exp (-a_k(\tau_k^\star - b_k) ) \big) -  1 &=& 0, \forall k \in \mathcal{K},
\end{eqnarray}
\end{subequations}
where $ E_k^{\text{In},\star} = \rho^\star  \  \big(\mathbf{h}_k^\dagger (\sum_{j=1}^K \mathbf{W}_j^\star + \mathbf{V}^\star) \mathbf{h}_k + \sigma_{k,S}^2\big) \geq \tau_k^\star$.

Now, the objective function has the \emph{log-concave} form and it can be solved given the sets $\bm{\mu}$ and $\bm{\epsilon}$. The iterative update of the vector sets can be carried out in the following way. Let us define the function $\mathcal{F}(\bm{\mu}, \bm{\epsilon}) = \big[ \epsilon_k \big(1+\exp (-a_k(\tau_k^\star - b_k) ) \big) - M_k , \ldots, \mu_k \big(1+\exp (-a_k(\tau_k^\star - b_k) ) \big) -  1 \big]$, $\forall k \in \mathcal{K}$. The next set of values of $\bm{\mu}$ and $\bm{\epsilon}$ can be updated by solving $\mathcal{F}(\bm{\mu}, \bm{\epsilon}) = \mathbf{0}$. Specifically, in the $q$-th iteration, we update them as:
\bea \label{update}
\bm{\mu}^{q+1}= \bm{\mu}^{q} + \varpi^q \mathbf{p}^q,  \ \bm{\epsilon}^{q+1} = \bm{\epsilon}^{q} + \varpi^q \mathbf{p}^q,
\eea
where $\mathbf{p}^q =  [\mathcal{F}'(\bm{\mu}, \bm{\epsilon})] \mathcal{F}(\bm{\mu}, \bm{\epsilon})$, $\mathcal{F}'(\bm{\mu}, \bm{\epsilon})$ is the Jacobian matrix of $\mathcal{F}(\bm{\mu}, \bm{\epsilon})$, $\varpi^q$ is the largest  $\varpi^l$ that satisfies $|| \mathcal{F}(\bm{\mu}^q + \varpi^l \mathbf{p}^q, \bm{\epsilon}^q + \varpi^l \mathbf{p}^q )|| \leq (1-t \varpi^l) ||\mathcal{F}(\bm{\mu}, \bm{\epsilon})||$, $l = 1,2,\ldots$, $0<\varpi^l<1$, and $0<t<1$ \cite{E. Boshkovska} \cite{Y. Jong}.

A two-loop algorithm is proposed for solving the problem. The outer loop gives $\bm{\mu}$ and $\bm{\epsilon}$ as the inputs of the inner loop, while the inner loop finds $\mathbf{W}_k^{\star}$ and $\mathbf{V}^{\star}$. Observe that in (\ref{p6_newobj}), there is a coupling variable $\frac{\tau_k}{\rho}$,  which  is convex with a given $\rho$. Therefore, in the inner loop, we have to perform a one-dimensional search for $\rho$ as well.  The detailed algorithm is formulated in Algorithm 1.
 \begin{algorithm} \label{alg1}
 \caption{Robust Precoding Design for EH Maximization Problem}
 \begin{algorithmic}[1]
 \\ \textbf{Input}: Minimum required data rate $R_k$ of SU $k$, noise power $\sigma_{k,S}^2$ and $\sigma_D^2$, channel uncertainty $\varphi_k^2$ and $\psi_n^2 $, maximum allowed interference power $P_{n,p}$ for PU $n$, maximum BS transmitted power $P_B$, and randomly generated estimated channel $\widehat{\mathbf{h}}_k$ and $\widehat{\mathbf{g}}_n$.
 \\ \textbf{Initialisation}:  Iteration number $q=0, p = 1$, initial value of $\rho$ as $\rho_{\text{start}}$, step $s$, end value   $\rho_{\text{end}}$, $\bm{\mu}^0$, and $\bm{\epsilon}^0$, loop stop criteria $m_{th}$.
 \\ \textbf{One-dimensional Search}:
  \FOR {$\rho$ =  $\rho_{\text{start}}$ :$s$: $\rho_{\text{end}}$}
  \REPEAT: \{Outer Loop\}
  \STATE Solve for the optimization problem $\mathbf{P}_7$: \{Inner Loop\}
\IF {($\mathbf{P}_7$ is feasible)}
 \STATE  Obtain $\mathbf{W}_k^q$ and $\mathbf{V}^q$.
 \ELSE \STATE Break from the outer loop.
 \ENDIF
 \STATE Update $\bm{\mu}^{q+1}$ and $\bm{\epsilon}^{q+1}$ according to (\ref{update}), then let $q= q + 1$.
 \UNTIL $\big|\mu_k^{q+1} \big\{ M_k - \epsilon_k^{q+1} \big(1+\exp (-a_k(\tau_k - b_k) ) \big) \big\} \big| < m_{th} $
 \STATE Calculate $E_{\text{sum}}^i = \sum_k E_k^{\text{Practical}}$, then let $i = i+1, q = 0. $
  \ENDFOR
\STATE Find the maximum value among all $E_{\text{sum}}^i$, and the precoding and energy matrix.
\\ \textbf{Output}: Use either of the methods to get the precoding vector $\mathbf{w}_k^{\text{opt}}$ and $\mathbf{V}^{\text{opt}}$.
 \end{algorithmic}
 \end{algorithm}

\subsection{Gaussian CSI Error Model}
In this section, we formulate the maximum harvested energy under the gaussian CSI error model formulated is as follows:
\begin{subequations}
\begin{eqnarray}
 \mathbf{P}_8 :  & &   \max_{\mathbf{W}_k  \in \mathbb{C}^{M \times M}, \mathbf{V} \in \mathbb{C}^{M \times M}, \rho}  \   \ \sum_{k=1}^K \  E_k^{\text{Practical}}  \label{p8_obj}\\
 \text{s.t.} & &   (\ref{p4c1}), (\ref{p4c3}), (\ref{p2c4}),(\ref{p2c5}), (\ref{p2c6}), (\ref{p2c7}).
\end{eqnarray}
\end{subequations}
We first simplify the objective function and then a new approximation will be formulated based on the \emph{Bernstein-type Inequality} \cite{prob1}\cite{prob2}. By involving a simple transformation, we arrive at:
\begin{subequations}
\begin{eqnarray}
 \mathbf{P}_9 :  & & \max_{\mathbf{W}_k , \mathbf{V},\rho }  \   \ \sum_{k=1}^{K} \  \mu_k \big\{ M_k - \epsilon_k \big(1+\exp (-a_k(\tau_k - b_k) ) \big) \big\}, \label{p9_obj}\\
 \text{s.t.} & & \text{Pr} (E_k^{\text{In}}  \geq \tau_k) \geq 1- \varpi, \   \ \forall \Delta  \mathbf{h}_k \sim \mathcal{CN}(0, \mathbf{H}_k) , \ \forall k \in \mathcal{K}, \label{p9c1}\\
 & & (\ref{p4c1}), (\ref{p4c3}), (\ref{p2c4}),(\ref{p2c5}), (\ref{p2c6}), (\ref{p2c7}).
\end{eqnarray}
\end{subequations}
Observe  however that the transformation from (\ref{p8_obj}) to (\ref{p9_obj}) and  (\ref{p9c1}) is not exactly equivalent. The equivalent form should let  $E_k^{\text{In}}  \geq \tau_k$ in (\ref{p9c1}). However, by setting $\varpi$ to be a very small value, the transformation can be valid and it is also consistent with our gaussian CSI error model. By applying the \emph{Bernstein-type Inequality I} \cite{prob1}, (\ref{p9c1}) becomes,
\begin{subequations}\label{p9c1_trans}
\begin{eqnarray}
\text{Tr} \big(\mathbf{H}_k^{1/2}\mathbf{\Sigma}  \mathbf{H}_k^{1/2} \big) - \sqrt{-2 \ln (\varpi)} \upsilon_{1k,s} + \ln (\varpi) \upsilon_{2k,s} + c_{k,s} \geq 0,  \\
c_{k,s} = \widehat{\mathbf{h}}_k^\dagger \mathbf{\Sigma}  \widehat{\mathbf{h}}_k + \sigma_{k,S}^2 - \frac{\tau_k}{\rho}, \\
\Bigg  | \Bigg |\begin{bmatrix}
 {\begin{array}{c}
   \text{vec}\big( \mathbf{H}_k^{1/2}\mathbf{\Sigma}  \mathbf{H}_k^{1/2} \big)  \\
   \sqrt{2} \mathbf{H}_k^{1/2}  \mathbf{\Sigma}  \widehat{\mathbf{h}}_k  \\
  \end{array} }
 \end{bmatrix} \Bigg | \Bigg| \leq \upsilon_{1k,s}, \\
\upsilon_{2k,s}  \mathbf{I} + \big(\mathbf{H}_k^{1/2}\mathbf{\Sigma}  \mathbf{H}_k^{1/2} \big) \succeq \mathbf{0}, \upsilon_{2k,s}  \geq 0, \ \forall k \in \mathcal{K},
\end{eqnarray}
\end{subequations}
where $\upsilon_{1k,s}$ and $\upsilon_{2k,s}$, $k \in \mathcal{K}$ are slack variables.

We also relax the problem by dropping the constraint that the rank of $\mathbf{W}_k$ must be 1, and the optimization problem becomes
\begin{subequations}
\begin{eqnarray}
 \mathbf{P}_{10} :   \max_{\substack{\mathbf{W}_k , \mathbf{V},\rho,  \{\upsilon_{1i,k}\}, \\ \{\upsilon_{2i,k}\},\{\upsilon_{1k,s}\},\{\upsilon_{2k,s}\}, \{\upsilon_{1,n}\}, \{\upsilon_{2,n}\} }} & & \   \ \sum_{k=1}^{K} \  \mu_k \big\{ M_k - \epsilon_k \big(1+\exp (-a_k(\tau_k - b_k) ) \big) \big\}, \label{p10_obj}\\
 \text{s.t.} \ \  & &   (\ref{p9c1_trans}), (\ref{p4c1_trans}), (\ref{p4c3_trans}), \\
 & & (\ref{p2c4}),(\ref{p2c5}), (\ref{p2c6}).
\end{eqnarray}
\end{subequations}
Still, the coupling variable in (\ref{p9c1_trans}) can be tackled by fixing $\rho$. A similar one-dimensional search for $\rho$, together with a two-loop algorithm can solve $\mathbf{P}_{10}$, the detailed step will be omitted here for space considerations.

\subsection{Complexity Analysis}
For the CBS power minimization problem under the bounded CSI model, $\mathbf{P}_3$ has $\frac{K (K+1)}{2}$ linear matrix inequality (LMI) constraints of size $(M+1)$ in (13) due to the higher decoding complexity.  Furthermore, we have $N$ LMI constraints of size $(M+1)$ in (15) and $K$ LMI constraints of size $(M+1)$ in (17). Additionally, in (12g), there are $(K+1)$ LMI constraints associated with size $M$, and a total of $\frac{K (K+1)}{2} + 2N + K +2$ linear constraints. Thus, according to \cite{F. Zhou2} and the reference therein, the total complexity becomes
\begin{eqnarray}
C_{\text{com}}^B = \ln (\tau ^{-1}) n \sqrt{\Psi_{\text{comp}}^1 } \bigg( (\frac{K (K+1)}{2} + N + 2K+1) [ (M+1)^3 + n (M+1)^2 ] \\ 
 + (K+1) (M^3 + n M^2) + \frac{K (K+1)}{2} + 2N + K +2 + n^2\bigg), \nonumber
\end{eqnarray}
where $n = \mathcal{O} \bigg((K+1)M^2+ N + K +\frac{K (K+1)}{2}  \bigg)	$, $\mathcal{O} $ is the big-O notation. Furthermore, we have $\Psi_{\text{comp}}^1 =  	(\frac{K (K+1)}{2} + N + 2K+1) M + 	K^2  + 4N + 3K + 4$, and $\tau$ is the accuracy of iteration.

Similarly, under Gaussian error model, there are  $3K (\frac{K+3}{2})+ 3N +2$ linear constraints, $\frac{K(K+1)}{2} + 2K + N + 1$ LMI of size $M$, and $\frac{K(K+1)}{2} + K + N$ second-order cone (SoC) constraints.   Thus, the complexity becomes:
\begin{eqnarray}
C_{\text{com}}^G = \ln (\tau^{-1}) n \sqrt{\Psi_{\text{comp}}^2 } \bigg(    (\frac{K(K+1)}{2} + 2K + N + 1) [ M^3 + n M^2 ] + 3K (\frac{K+3}{2})+ 3N +2 \\ \nonumber + (\frac{K(K+1)}{2} + K + N) [(M^2+M+1)^2 ] + n^2 \bigg)
\end{eqnarray}
Where $\Psi_{\text{comp}}^2  = 3 K^2 + 10 K + 6N + 3$.

For the maximum harvested energy problem, with bounded channel model,
since the difference with that of power minimization problem is that a maximum of $T_{\text{max}}$ number of iterations will be performed for one-dimensional search. Hence, the complexity is $T_{\text{max}} C_{\text{com}}^B$. With Gaussian error model, the complexity is $T_{\text{max}}^{'} C_{\text{com}}^G$, correspondingly, $T_{\text{max}}^{'}$ is the number of unitary search.

\section{Simulation Results}
In this section, we present our simulation results for characterizing  the performance of  the  proposed robust beamforming  conceived with NOMA under both the bounded and the gaussian CSI  estimation error models. Unless otherwise stated, the parameters are chosen as in Table.\ref{table1}.
\begin{table}[ht]
\centering
\begin{tabular}{ | c  | c | }
\hline
Parameters & Values \\
\hline
Number of SUs and PUs & $K = 3$, $N = 2$\\
\hline
Noise  powers & $\sigma_{k,S}^2 = 0.1$, $\sigma_D^2 = 0.01$ \\
\hline
Minimum required EH power & $P_{k,s} = 0.01$ Watt \\
\hline
Maximum tolerable interference of PUs & $P_{n,p} = -18$ dBm \\
\hline
Estimated channel gains & $\hat{\mathbf{h}}_k \sim \mathcal{CN} (0, 0.8 \mathbf{I})$ , $\hat{\mathbf{g}}_n \sim \mathcal{CN} (0, 0.1 \mathbf{I})$\\
\hline
Outage probability threshold & $\xi_k = \xi_{k,s} = \xi_{n,p} = 0.05$\\
\hline
Gaussian CSI estimation &  $\varpi_{k}^2 = 0.001$, $\varpi_{n}^2 = 0.0001$ \cite{F. Zhou2}\\
\hline
Non-linear EH model & $M_k = 24$ mW, $a_k = 150$, and $b_k = 0.014$  \cite{Non_EH_para}\\
\hline
\end{tabular}
\caption{Simulation Parameters}
\label{table1}
\end{table}

To achieve a fair comparison between the two channel estimation error models. If the covariance matrices of the channel estimation error vector $\Delta \mathbf{h}_k$ and $\Delta \mathbf{g}_n$ under the gaussian model are $\varpi_{k}^2 \mathbf{I}$ and $\varpi_{n}^2 \mathbf{I}$, respectively, then the bounded CSI radius under the worst-case scenario of $\varphi_k$ and $\psi_n$ should be \cite{Y. Wang}
\begin{subequations}\label{W_to_P_relation}
\begin{eqnarray}
\varphi_k = \sqrt{\frac{\varpi_k^2 F_{2M}^{-1}(1-\xi_k)}{2}},
\ \psi_n = \sqrt{\frac{\varpi_n^2 F_{2M}^{-1}(1-\xi_{n,p})}{2}},
\end{eqnarray}
\end{subequations}
where $F_{2M}^{-1}(\cdot)$ represents the complimentary cumulative distribution function (CCDF) of the Chi-square distribution with $2M$ degrees of freedom.
\subsection{Power Minimization Problem}
\begin{figure}[h]
\centering
\includegraphics[width=3.8in]{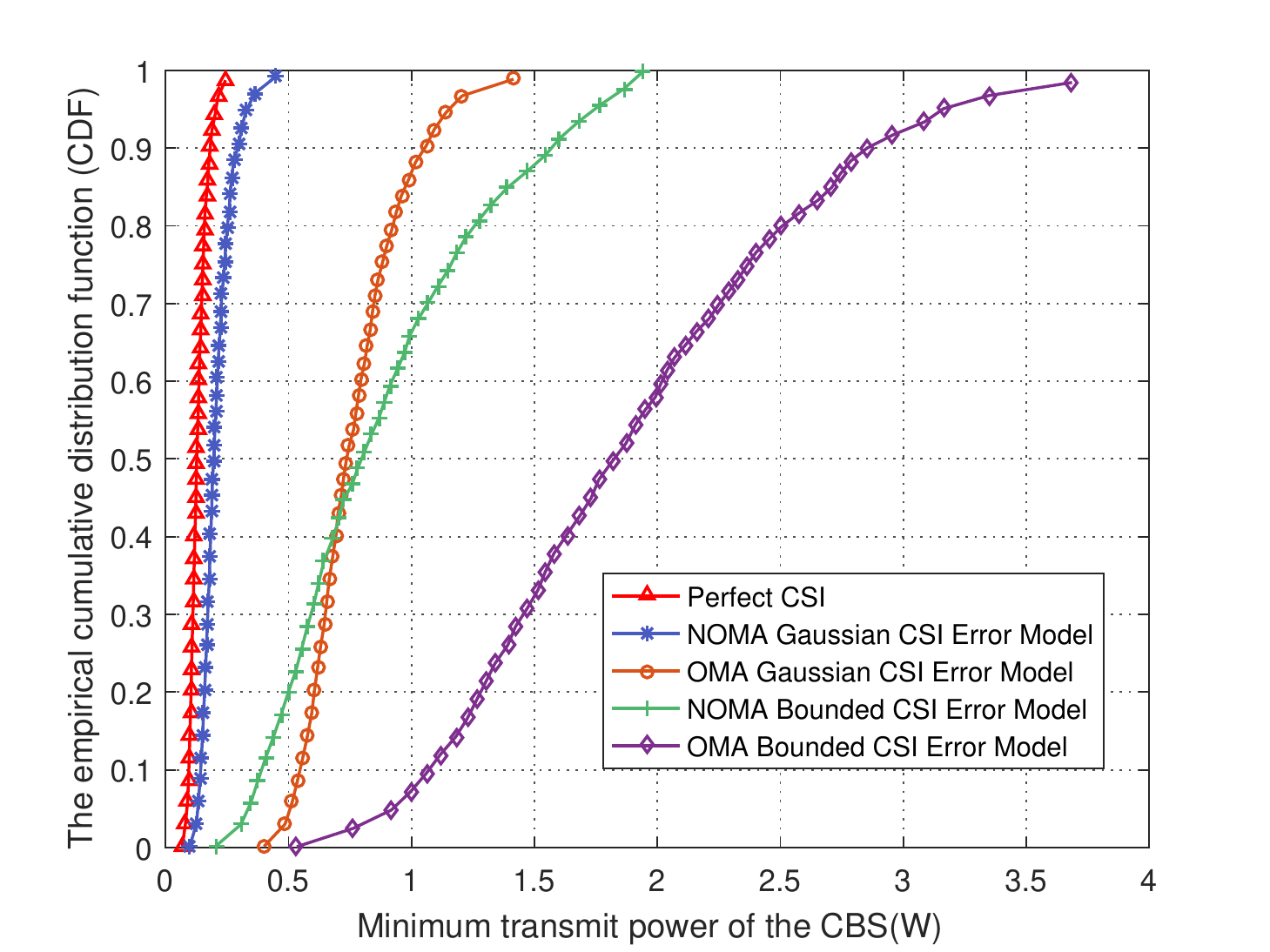}
\caption{The empirical CDF of the minimum transmit power of the CBS under different channel conditions. CBS antenna number $M = 10$, $P_B = 2$ Watts, $R_{\text{min}} = 1$ bit/s/Hz.}
\label{result1}
\end{figure}

Fig. \ref{result1} shows the empirical CDFs of the minimum transmit power of the CBS for both the imperfect CSI estimation error models. The maximum power $P_B$ is set to 2 Watts. For comparison, we also include the case of OMA, since it represents the traditional access technology. Observe that in order to reduce the inter-user interference, each OMA user relies exclusively on a single time slot. Thus, a total of $K$ time slots are required instead of a single one in our scheme. To make a fair comparison, each SU's achievable data rate should be averaged over all K time slots, which becomes $R_k^{\text{OMA}} = \frac{1}{K} \log_2 (1 + \text{SINR}_k^{\text{OMA}})$. Reduced interference is achieved at the cost of a lower spectral and energy efficiency.  We also observe that under both channel error models, the performance of NOMA is better than that of OMA. This is because for OMA, the lower spectral efficiency makes the SU data rate requirement harder to be satisfied. Hence the CBS has to apply a higher transmission power to compensate for that, which leads to a much higher energy consumption.  Fig. \ref{result1} is generated from 1,000 independent realizations of different channel conditions.  As expected, the performance under perfect CSI is the best, since no additional power is used to compensate for the channel uncertainties. Furthermore, in both the OMA and NOMA schemes, the performance under the gaussian CSI channel estimation is better than that under the bounded CSI channel estimations, as bounded CSI represents the worst-case scenario. Observe that the minimum power in the OMA bounded CSI is over 2 Watts since we only limit the power of each time slot to 2 Watts and it is very likely that the total power over $K$ slots will beyond that limit.

\begin{figure}
\centering
\includegraphics[width=3.8in]{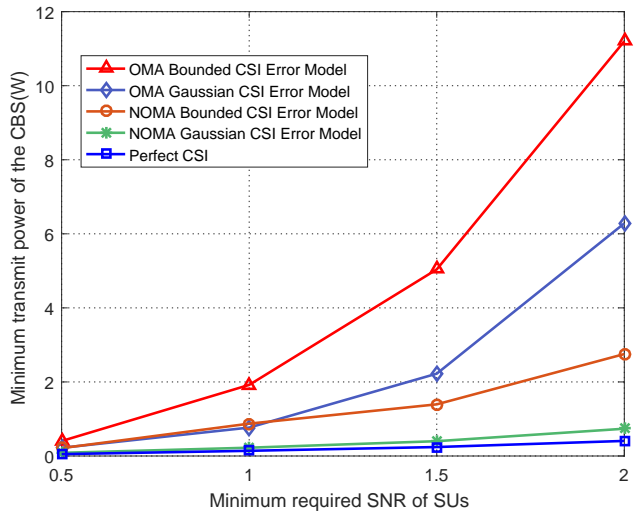}
\caption{The minimum transmit power of the CBS vs. the required SNR of SUs for $M=10$, $P_B = 8$ Watts.}
\label{result2}
\end{figure}

 Fig. \ref{result2} shows  the minimum transmit power of the CBS as a function of the minimum required SNR of SUs, $\gamma_{k,\text{min}}$. As the SNR increases, the power increases under all CSI cases. Also, perfect CSI requires  the least power,  followed by NOMA relying on the gaussian CSI error model, NOMA in the bounded CSI model, OMA gaussian CSI model, and OMA bounded CSI model. Besides, compared to OMA, the CBS power in NOMA grows more slowly. In the parameter setting,  $\gamma_{k,\text{min}}$ plays a more important role in the constraints. For $\gamma_{k,\text{min}} = 2 $ in the NOMA case, the equivalent SNR for OMA will be 26. Thus, the gap between OMA and NOMA further increases with the required SNR.

\begin{figure*}
\centering
\begin{subfigure}{.475\columnwidth}
\centering
\includegraphics[width=3.2in]{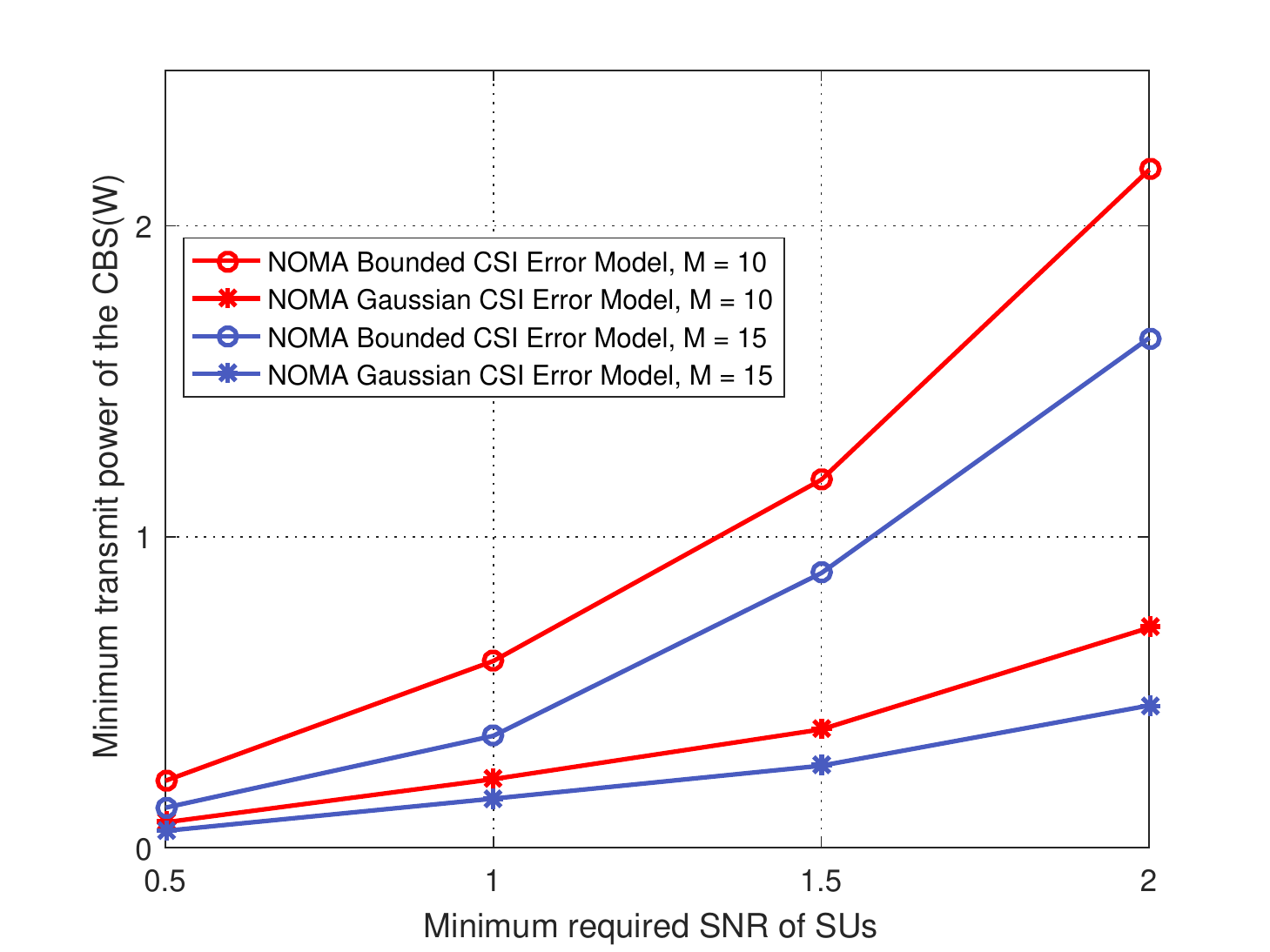}
\caption{}
\label{result3a}
\end{subfigure}
\hfill
\begin{subfigure}{.475\columnwidth}
\centering
\includegraphics[width=3.2in]{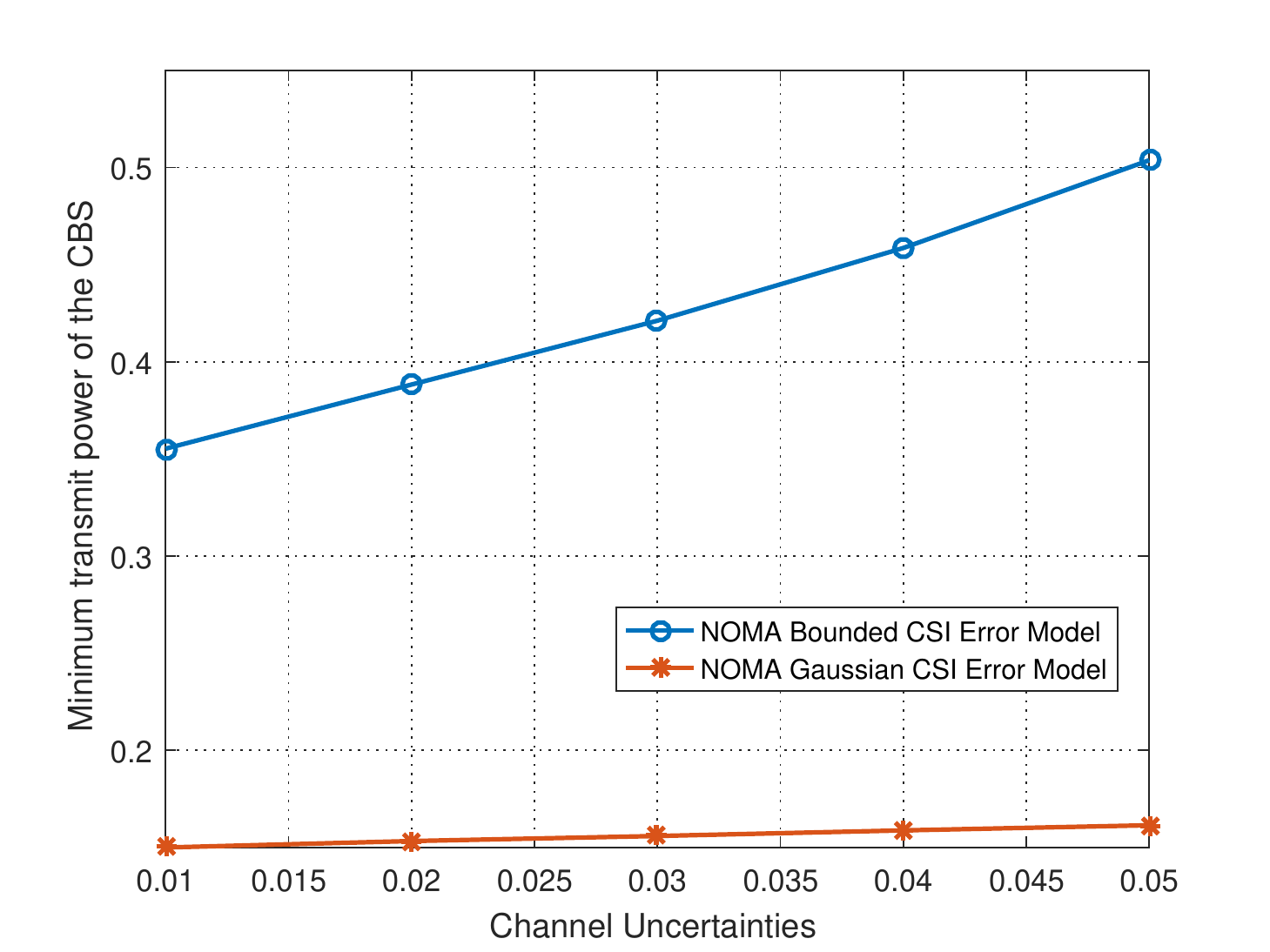}
\caption{}
\label{result3b}
\end{subfigure}
\caption{(a) Impact of the number of CBS antennas on the minimum transmitted power required in two imperfect CSI scenarios. (b) Impact of channel uncertainties $\psi_n$ and $\varphi_k$ on the overall minimum transmit power of the CBS, $M =15$, $R_{\text{min}} = 1 $ bit/s/Hz, $P_B = 8$ Watts.}
\label{result3}
\end{figure*}

The impact of the CBS antenna number is illustrated in Fig. \ref{result3}(a), where the performance with different CBS antenna numbers and channel uncertainties are plotted. Specifically,  Fig. \ref{result3}(a) illustrates how the number of antennas affects the overall performance. The power required increases, when the SNR of SUs grows, regardless of how many antennas are mounted at the CBS. It is also observed that the minimum power required decreases when the number of antennas increases, since a larger number of antennas results in a higher degree of freedom (DoF). Besides, we also notice that the performance under the gaussian error model is better than that under the bounded channel error case. In Fig. \ref{result3}(b), the impact of channel uncertainties is illustrated. We set $\psi_n^2 = \varphi_k^2 = [0.01:0.05]$, the corresponding covariance matrices in gaussian CSI estimation error scenario also change according to (\ref{W_to_P_relation}). Clearly, channel estimation error affects the bounded CSI scenario the most, since under worst-case CSI, the channel estimation error channel becomes worse, thus it needs more power to meet the data rate constraints. Nevertheless, the channel estimation error does not have much impact on the gaussian channel estimation error scenario.

\subsection{Energy Harvesting Maximization Problem}
\begin{figure}
\centering
\includegraphics[width=3.6in]{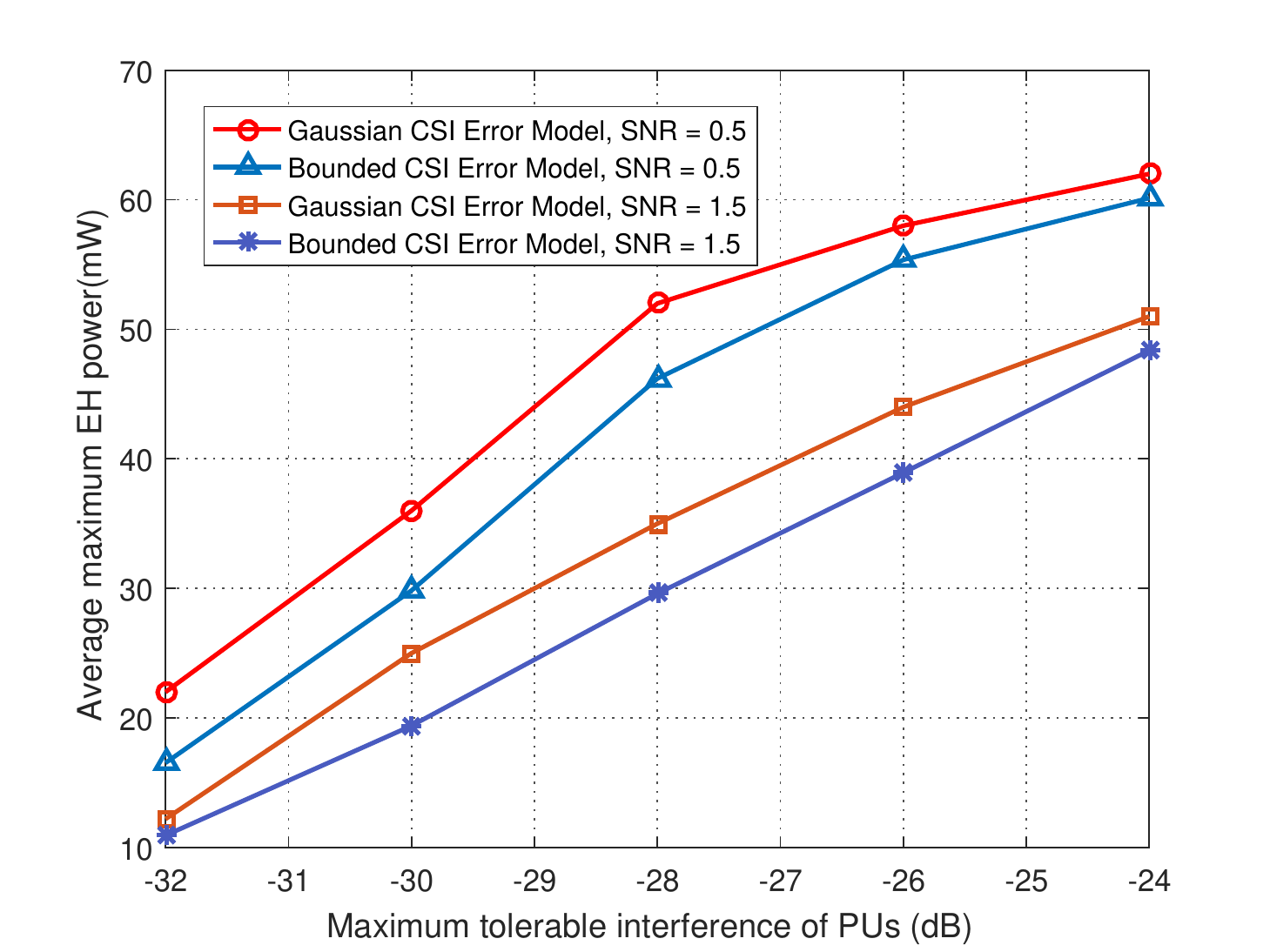}
\caption{Average maximum EH power  under different  interferences tolerated by the PUs, $M=10$.}
\label{result4}
\end{figure}

In this subsection, we present results for the maximum EH as our objective function. The CBS power is $P_B = 2$ Watts. Fig. \ref{result4} characterizes the average maximum EH power vs. the interference tolerated by the to PUs. One can observe that the energy harvested monotonically increases, when the maximum interference tolerated by the PUs grows, where a higher $P_{n,p}$ allows for a larger transmission power, leading to the increase of the harvested energy. Additionally, we can see that under the gaussian channel estimation error, the performance is better than that under the bounded channel estimation error case. When the channel conditions are better, less power is required for satisfying the  data rate requirements. Hence more power can be reserved for EH.  This also explains that when the required SNR is low, a high EH power can be achieved.

\begin{figure}
\centering
\includegraphics[width=3.8in]{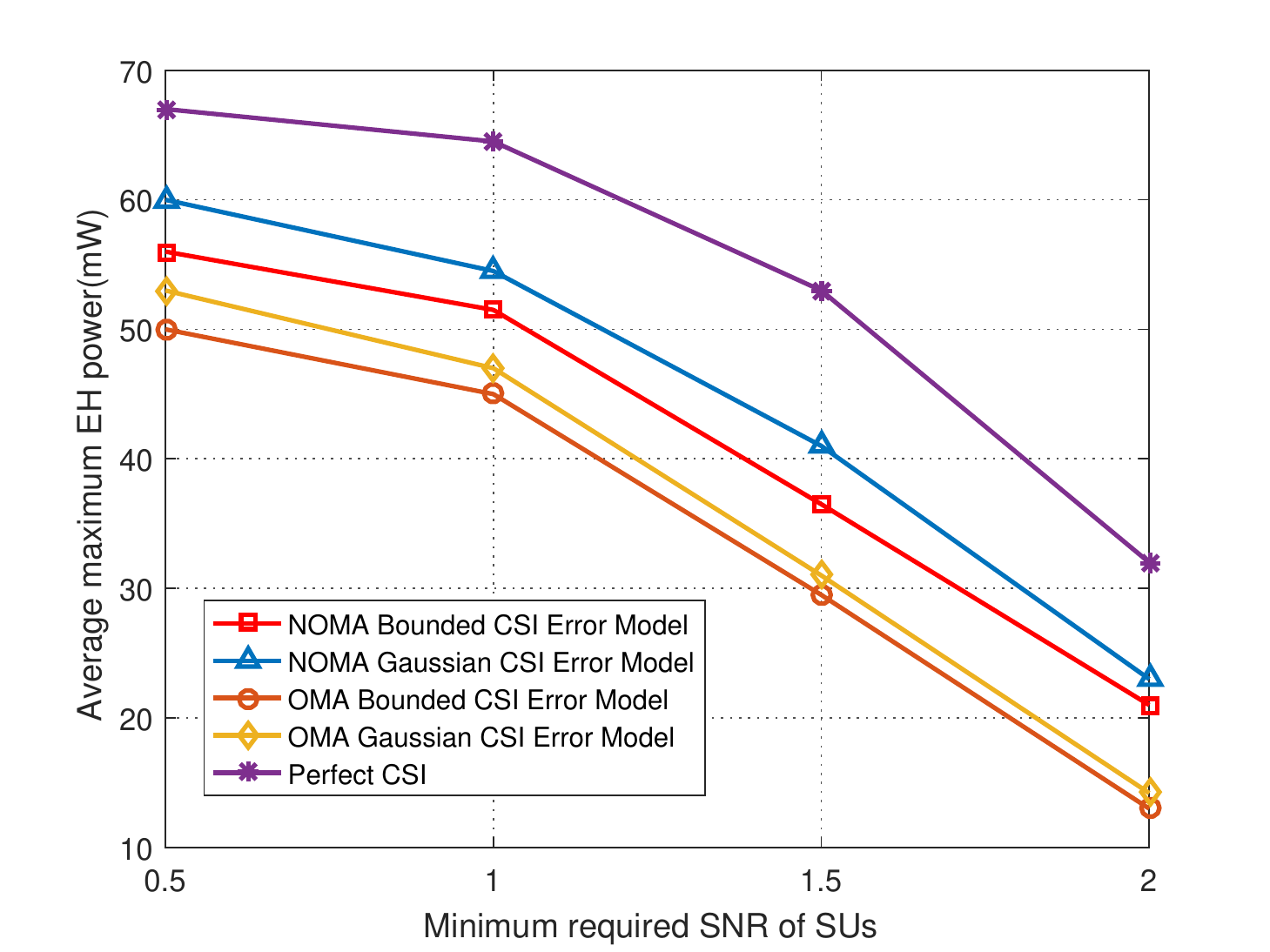}
\caption{Average maximum EH power vs. the minimum SNR required by the SUs, $M=10$.}
\label{result5}
\end{figure}

The impact of minimum SNRs required by the SUs is illustrated in Fig. \ref{result5}. The number of CBS antennas is $M = 10$ and the interference threshold $P_{n,p}$ is set to -24 dBm. We also list the results for the OMA cases. As expected, the average maximum EH power decreases, when the required SNR increases. Similar observations show that under perfect CSI, the performance is the best, while the OMA bounded CSI estimation scenario is the worst. Moreover, we can see that the maximum EH power decreases significantly when the SNR grows. This is because more power has to be used for information detection, which leaves less power for energy harvesting.

\begin{figure}[h]
\centering
\includegraphics[width=3.8in]{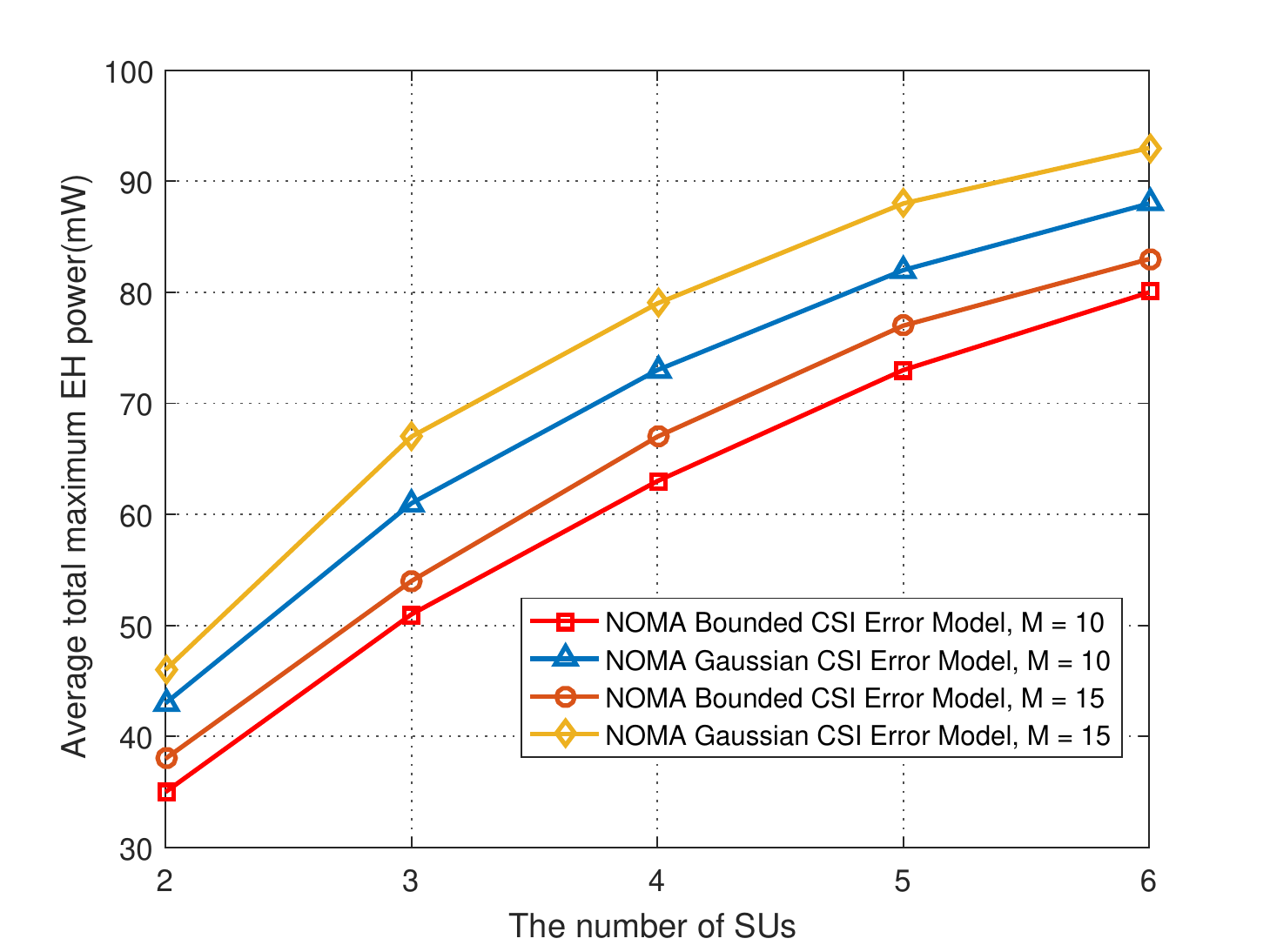}
\caption{Average total EH power vs. the number of SUs for $P_{n,p} = -24$ dBm, $r_{\text{min}} = 1$ bit/s/Hz.}
\label{result6}
\end{figure}

Fig. \ref{result6} shows the average total EH power vs. the number of SUs. It can be observed that the total EH power grows, when the number of SUs increases, since more nodes participate in the harvesting process. Additionally, we can see that when the number of antennas is higher, more EH power can be achieved. This is because more antennas give a higher system DoF, therefore less power is sufficient for information detection.

\section{Conclusions}
In this paper, we considered MISO-NOMA CR-aided SWIPT under both the bounded and the Gaussian CSI estimation error model. To make the energy harvesting investigations more realistic, a non-linear EH model was applied. Robust beamforming and power splitting control were jointly designed for achieving the  minimum transmission power and  maximum EH. We transformed the non-convex minimum transmission power optimization problems into a convex form while applying a one-dimensional search algorithm to solve the maximum EH problem. Our simulation results showed that the performance achieved by using NOMA is better than that obtained by using the traditional OMA. Furthermore, a performance gain can be obtained under the gaussian CSI estimation error model over the bounded CSI error model. As for future research directions, the system model can be generalized to account for more use cases, for example, considering the physical layer security and the interference arising from multipoe cells. Additionally, for the Gaussian CSI error model, the rank of the solution is not fully characterized in this work.
\appendices
\section{{Proof of Theorem 3.2}}
\begin{IEEEproof}
To prove the Lemma, we first consider the Karush-Kuhn-Tucker (KKT) conditions of $\mathbf{P}_3$. Specifically, after some simple algebraic manipulations, (\ref{newC1}) can be rewritten as
\bea
\begin{bmatrix}
  {\begin{array}{cc}
   \alpha_{i,k} \mathbf{I}  & \mathbf{0} \\
   \mathbf{0}  & t_{i,k} \\
  \end{array} }
  \end{bmatrix}   +
  \begin{bmatrix}
  {\begin{array}{c}
   \mathbf{I}  \\
   \widehat{\mathbf{h}}_i^\dagger \\
  \end{array} }
  \end{bmatrix}  \mathbf{C}_k
   \begin{bmatrix}
  {\begin{array}{cc}
   \mathbf{I} & \widehat{\mathbf{h}}_i \\
  \end{array} }
  \end{bmatrix}   +
  \begin{bmatrix}
  {\begin{array}{cc}
   -\gamma_{k,\text{min}} \sum_{j=1}^{k-1} \mathbf{W}_j  & \mathbf{0} \\
   \mathbf{0}  & 0 \\
  \end{array} }
  \end{bmatrix}  \succeq \mathbf{0}, \\ \nonumber
  \ \forall k \in \mathcal{K}, \ i = \{k, k+1, \ldots, K\},
\eea
where we have $t_{i,k}= -\alpha_{i,k} \varphi_k^2 - \gamma_{k,\text{min}} \big( \sigma_{k,S}^2 + \frac{\sigma_D^2}{(1-\rho)}\big)$.

Similarly, (\ref{newC3}) and \ref{newC2} can be rewritten as
\bea
\begin{bmatrix}
  {\begin{array}{cc}
   \beta_{n} \mathbf{I}  & \mathbf{0} \\
   \mathbf{0}  & -\beta_{n} \psi_n^2 + P_{n,p} \\
  \end{array} }
  \end{bmatrix}   -
  \begin{bmatrix}
  {\begin{array}{c}
   \mathbf{I}  \\
   \widehat{\mathbf{g}}_n^\dagger \\
  \end{array} }
  \end{bmatrix} \mathbf{\Sigma}
   \begin{bmatrix}
  {\begin{array}{cc}
   \mathbf{I} & \widehat{\mathbf{g}}_n \\
  \end{array} }
  \end{bmatrix}  \succeq \mathbf{0}, \ \forall n \in \mathcal{N},
\eea

and

\bea
\begin{bmatrix}
  {\begin{array}{cc}
   \theta_{k} \mathbf{I}  & \mathbf{0} \\
   \mathbf{0}  & m_{k} \\
  \end{array} }
  \end{bmatrix}   +
  \begin{bmatrix}
  {\begin{array}{c}
   \mathbf{I}  \\
   \widehat{\mathbf{h}}_k^\dagger \\
  \end{array} }
  \end{bmatrix}  \mathbf{\Sigma}
   \begin{bmatrix}
  {\begin{array}{cc}
   \mathbf{I} & \widehat{\mathbf{h}}_k \\
  \end{array} }
  \end{bmatrix}   \succeq \mathbf{0}, \ \forall k \in \mathcal{K},
\eea
respectively, where $m_{k} = -\theta_{k} \varphi_k^2 +  \sigma_{k,S}^2 - \frac{\tau_k}{\rho}$.

For notational simplicity, we let $\mathbf{X}_i =  \begin{bmatrix}
  {\begin{array}{cc}
   \mathbf{I} & \widehat{\mathbf{h}}_i \\
  \end{array} }
  \end{bmatrix}$ and $\mathbf{Y}_n =  \begin{bmatrix}
  {\begin{array}{cc}
   \mathbf{I} & \widehat{\mathbf{g}}_n \\
  \end{array} }
  \end{bmatrix}$. Furthermore, we denote $\mathbf{A}_{i,k} \in \mathbb{C}_{+}^{(M+1) \times (M+1)}$, $\mathbf{B}_{k} \in \mathbb{C}_{+}^{(M+1) \times (M+1)}$, $\mathbf{D}_{n} \in \mathbb{C}_{+}^{(M+1) \times (M+1)}$, $z \in \mathbb{R}_{+}$, and $\mathbf{E}_{k} \in \mathbb{C}_{+}^{(M) \times (M)}$ as the KKT multiplier. Then the Lagrange dual function $\mathbf{\mathcal{L}}$ can be expressed as
\bea
& & \mathbf{\mathcal{L}}(\mathbf{W}_k,\mathbf{V}, \mathbf{A}_{i,k},\mathbf{B}_{k}, \mathbf{D}_{n}, z, \kappa) = \text{Tr}(\mathbf{\Sigma}) - \sum_{i,k} \text{Tr} (\mathbf{A}_{i,k} \mathbf{X}_i^\dagger \mathbf{C}_k \mathbf{X}_i) - \sum_{i,k} \text{Tr} (\mathbf{A}_{i,k} \mathbf{M}_k )  \\ \nonumber& & +  \sum_{n} \text{Tr} (\mathbf{D}_{n} \mathbf{Y}_n^\dagger \mathbf{\Sigma} \mathbf{Y}_n) - \sum_{k} \text{Tr} \big(\mathbf{B}_{k} \mathbf{X}_k^\dagger \mathbf{\Sigma} \mathbf{X}_k \big) + z \big( \text{Tr}(\mathbf{\Sigma}) - P_B \big) - \sum_{k} \text{Tr} (\mathbf{E}_k \mathbf{W}_k) + \kappa,
\eea
where $\mathbf{M}_k =  \begin{bmatrix}
  {\begin{array}{cc}
   -\gamma_{k,\text{min}} \sum_{j=1}^{k-1} \mathbf{W}_j  & \mathbf{0} \\
   \mathbf{0}  & 0 \\
  \end{array} }
  \end{bmatrix} $ and $\kappa$ are  the terms independent of $\mathbf{W}_k$.  Taking  the partial derivative of the dual function with respect to $\mathbf{W}_k$, we have

\bea \label{dual}
\frac{\partial \mathcal{L} }{\partial \mathbf{W}_k} = \mathbf{I} - \sum_i \mathbf{X}_i \mathbf{A}_{i,k} \mathbf{X}_i^\dagger + \gamma_{k,\text{min}} \sum_i \sum_{j=1}^{k-1}  \mathbf{X}_i \mathbf{A}_{i,j} \mathbf{X}_i^\dagger + \sum_i \gamma_{k,\text{min}} \sum_{j=k+1}^{K} \mathbf{A}_{i,j} \\ \nonumber + \sum_{n} \mathbf{Y}_n \mathbf{D}_n \mathbf{Y}_n^\dagger - \sum_{k} \mathbf{X}_k \mathbf{B}_k \mathbf{X}_k^\dagger + z \mathbf{I} - \mathbf{E}_k = \mathbf{0}.
\eea
Additionally, the dual problem has to satisfy the completeness slackness
\begin{subequations}
\begin{eqnarray}
 \big(\begin{bmatrix}
  {\begin{array}{cc}
   \alpha_{i,k} \mathbf{I}  & \mathbf{0} \\
   \mathbf{0}  & t_{i,k} \\
  \end{array} }
  \end{bmatrix} +  \mathbf{X}_i^\dagger  \mathbf{C}_k \mathbf{X}_i + \mathbf{M}_k \big) \mathbf{A}_{i,k} = \mathbf{0}, \label{dual1}\\
\mathbf{E}_k \mathbf{W}_k =  \mathbf{0}, \forall k \in \mathcal{K}, i = \{k+1, \ldots, K\}, \forall n \in \mathcal{N} \label{slackness}.
\end{eqnarray}
\end{subequations}
Right-multiplying  $\mathbf{W}_k$ with (\ref{dual}), and substituting (\ref{slackness}), we arrive at:
\bea
(\sum_i \mathbf{X}_i  \mathbf{A}_{i,k} \mathbf{X}_i^\dagger + \sum_{k} \mathbf{X}_k \mathbf{B}_k \mathbf{X}_k^\dagger  )\mathbf{W}_k =   \big [(1+z) \mathbf{I} +  \gamma_{k,\text{min}} \sum_i \sum_{j=1}^{k-1}  \mathbf{X}_i \mathbf{A}_{i,j} \mathbf{X}_i^\dagger \\ \nonumber + \gamma_{k,\text{min}} \sum_i \sum_{j=k+1}^{K} \mathbf{A}_{i,j} + \sum_{n} \mathbf{Y}_n \mathbf{D}_n \mathbf{Y}_n^\dagger \big] \mathbf{W}_k.
\eea
Since all the KKT multipliers are positive numbers or positive semi-definite matrices, we can readily verify $\big\{(1+z) \mathbf{I} +  \gamma_{k,\text{min}} \sum_i \sum_{j=1}^{k-1}  \mathbf{X}_i \mathbf{A}_{i,j} \mathbf{X}_i^\dagger + \gamma_{k,\text{min}} \sum_i \sum_{j=k+1}^{K} \mathbf{A}_{i,j} + \sum_{n} \mathbf{Y}_n \mathbf{D}_n \mathbf{Y}_n^\dagger \big \} \succeq \mathbf{0}$. Thus it is non-singular. Left-multiplying a non-singular matrix with  $\mathbf{W}_k$ does not change  the  rank of  $\mathbf{W}_k$ . Therefore, we have
\bea
\text{Rank} (\mathbf{W}_k) & =& \text{Rank} \big ((\sum_i \mathbf{X}_i  \mathbf{A}_{i,k} \mathbf{X}_i^\dagger + \sum_{k} \mathbf{X}_k \mathbf{B}_k \mathbf{X}_k^\dagger  )\mathbf{W}_k  \big ) \\ \nonumber
&  \leq  & \min \big\{ \text{Rank}\big (\sum_i \mathbf{X}_i  \mathbf{A}_{i,k} \mathbf{X}_i^\dagger + \sum_{k} \mathbf{X}_k \mathbf{B}_k \mathbf{X}_k^\dagger   \big ), \text{Rank} (\mathbf{W}_k)\big\}.
\eea
Next, we show that the rank of $(\sum_i \mathbf{X}_i  \mathbf{A}_{i,k} \mathbf{X}_i^\dagger)$ is 1. By summing (\ref{dual1}) in terms of the index $i$, then left-multiplying $ \begin{bmatrix}
  {\begin{array}{cc}
   \mathbf{I}_M  & \mathbf{0}
  \end{array} }
  \end{bmatrix}$ and right-multiplying $\mathbf{X}_i^\dagger$, we have
\bea
\sum_i \alpha_{i,k} \mathbf{X}_i \mathbf{A}_{i,k} \mathbf{X}_i^\dagger - \sum_i \alpha_{i,k}  \begin{bmatrix}
  {\begin{array}{cc}
   \mathbf{0}_M  & \mathbf{h}_i
  \end{array} }
  \end{bmatrix} \mathbf{A}_{i,k} \mathbf{X}_i^\dagger + \sum_i \mathbf{C}_k \mathbf{X}_i \mathbf{A}_{i,k} \mathbf{X}_i^\dagger \\ \nonumber \
  +  \sum_i (-\gamma_{k,\text{min}} \sum_{j=1}^{k-1} \mathbf{W}_j ) \mathbf{X}_i \mathbf{A}_{i,k} \mathbf{X}_i^\dagger - \sum_i (-\gamma_{k,\text{min}} \sum_{j=1}^{k-1} \mathbf{W}_j ) \begin{bmatrix}
  {\begin{array}{cc}
   \mathbf{0}_M  & \mathbf{h}_i
  \end{array} }
  \end{bmatrix} \mathbf{A}_{i,k} \mathbf{X}_i^\dagger = \mathbf{0}.
\eea
After a simple transformation, we have
\bea
\sum_i \big( \alpha_{i,k} \mathbf{I}+ \mathbf{C}_k -\gamma_{k,\text{min}} \sum_{j=1}^{k-1} \mathbf{W}_j \big) \mathbf{X}_i \mathbf{A}_{i,k} \mathbf{X}_i^\dagger = \sum_i ( \alpha_{i,k}\mathbf{I} -\gamma_{k,\text{min}} \sum_{j=1}^{k-1} \mathbf{W}_j) \begin{bmatrix}
  {\begin{array}{cc}
   \mathbf{0}_M  & \mathbf{h}_i
  \end{array} }
  \end{bmatrix} \mathbf{A}_{i,k} \mathbf{X}_i^\dagger.
\eea
From the fact that (\ref{newC1}) is a positive semidefinite matrix, $\big( \alpha_{i,k} \mathbf{I}+ \mathbf{C}_k -\gamma_{k,\text{min}} \sum_{j=1}^{k-1} \mathbf{W}_j \big)$ would be a non-singular matrix, thus the rank of the left term of the above equation is the same as $ \sum_i \mathbf{X}_i \mathbf{A}_{i,k} \mathbf{X}_i^\dagger $. Also, it is easy to verify that the right term has a rank  1.

Similarly, we can prove that $\text{Rank}(\sum_{n} \mathbf{Y}_n \mathbf{D}_n \mathbf{Y}_n^\dagger\ ) = 1$. Then, the following equation holds:
\bea
\text{Rank}\big (\sum_i \mathbf{X}_i  \mathbf{A}_{i,k} \mathbf{X}_i^\dagger + \sum_{k} \mathbf{X}_k \mathbf{B}_k \mathbf{X}_k^\dagger   \big ) &\leq& \text{Rank} (\sum_i \mathbf{X}_i \mathbf{A}_{i,k} \mathbf{X}_i^\dagger ) + \text{Rank}(\sum_{n} \mathbf{Y}_n \mathbf{D}_n \mathbf{Y}_n^\dagger\ ) \nonumber \\
& \leq & 2,
\eea
which proves the theorem.
\end{IEEEproof}

\end{document}